\documentclass[10pt,journal,compsoc]{IEEEtran}
\usepackage{mathrsfs}

\usepackage{enumerate}
\usepackage{float}
\usepackage{hyperref}       
\usepackage{url}            
\usepackage{booktabs}       
\usepackage{amsfonts}       
\usepackage{nicefrac}       
\usepackage{microtype}      
\usepackage{times}
\usepackage{xcolor}
\usepackage{soul}

\usepackage{tikz}

\usepackage{framed}
\usepackage{lineno,hyperref}
\usepackage{enumerate}
\usepackage{amsfonts}       
\usepackage{graphicx}  
\usepackage{subfigure}
\usepackage{multirow}
\usepackage{amsmath}
\usepackage{float}
\usepackage{color}
\usepackage[noend]{algpseudocode}

\ifCLASSOPTIONcompsoc
  \usepackage[nocompress]{cite}
\else
  \usepackage{cite}
\fi
\ifCLASSINFOpdf
\else
\fi
\begin{document}
%
\title{Edge Deep Learning Model Protection via Neuron Authorization}
%
%
%
%

\author{Jinyin~Chen,
        Haibin Zheng,
        Tao Liu,
        Rongchang Li,
        Yao Cheng,
        Xuhong Zhang,
        Shouling Ji
\IEEEcompsocitemizethanks{\IEEEcompsocthanksitem This research was supported by the National Natural Science Foundation of China (No. 62072406), 
the National Key Laboratory of Science and Technology on Information System Security (No. 61421110502), 
the National Natural Science Foundation of China (No. U21B2001), 
the Key R\&D Programs of Zhejiang Province (No. 2022C01018). 
\IEEEcompsocthanksitem J.~Chen and H. Zheng are with the Institute of Cyberspace Security and the College of Information Engineering at Zhejiang University of Technology, Hangzhou, 310023, China (e-mail: chenjinyin@zjut.edu.cn, haibinzheng320@gmail.com).
\IEEEcompsocthanksitem T.~Liu, R.~Li are with the College of Information Engineering at Zhejiang University of Technology, Hangzhou, 310023, China (e-mail: leonliu022@163.com, lrcgnn@163.com).
\IEEEcompsocthanksitem Y.~Cheng is with the Huawei International, Singapore (e-mail: c.candyao@gmail.com).
\IEEEcompsocthanksitem X.~Zhang is with the College of Control Science and Engineering, Zhejiang University, Hangzhou, 310007, China (e-mail: zhangxuhong@zju.edu.cn).
\IEEEcompsocthanksitem S.~Ji is with the College of Computer Science and Technology at Zhejiang University, Hangzhou 310007, China (e-mail: sji@zju.edu.cn).
\IEEEcompsocthanksitem  Corresponding author: Haibin Zheng, e-mail: haibinzheng320@gmail.com.
}
\thanks{Manuscript received xx xx, 2022; revised xx xx, xxxx.}}

%
%

\markboth{XXXXXXXXXXXXXXX,~Vol.~XX, No.~XX, XXXXXXX XXXX}%
{Shell \MakeLowercase{\textit{et al.}}: EdgePro: Edge Deep Learning Model Protection via Neuron Authorization}
%



\IEEEtitleabstractindextext{%
\begin{abstract}
With the development of deep learning processors and accelerators, deep learning models have been widely deployed on edge devices as part of the Internet of Things.
Edge device models are generally considered as valuable intellectual properties that are worth for careful protection.
Unfortunately, these models have a great risk of being stolen or illegally copied.
The existing model protections using encryption algorithms are suffered from high computation overhead which is not practical due to the limited computing capacity on edge devices.
In this work, we propose a light-weight, practical, and general \underline{\emph{Edge}} device model \emph{\underline{Pro}tection} method at neuron level, denoted as EdgePro.
Specifically, we select several neurons as authorization neurons and set their activation values to locking values and scale the neuron outputs as the ``passwords'' during training.
EdgePro protects the model by ensuring it can only work correctly when the ``passwords'' are met, at the cost of encrypting and storing the information of the ``passwords'' instead of the whole model.
Extensive experimental results indicate that EdgePro can work well on the task of protecting on datasets with different modes.
The inference time increase of EdgePro is only 60\% of state-of-the-art methods, and the accuracy loss is less than 1\%.
Additionally, EdgePro is robust against adaptive attacks including fine-tuning and pruning, which makes it more practical in real-world applications.
EdgePro is also open sourced to facilitate future research: https://github.com/Leon022/EdgePro.
\end{abstract}

}

\maketitle

\IEEEdisplaynontitleabstractindextext

%
\IEEEpeerreviewmaketitle

\IEEEraisesectionheading{\section{Introduction}\label{sec:introduction}}
\IEEEPARstart{T}{he} wide use of deep learning models in Internet of Things (IoT) has greatly facilitated humans in the fields of smart city, intelligent medical treatment, and industrial manufacturing~\cite{sisinni2018industrial,gulati2019dilse,khalil2021deep,chehab2020lp}.
Deploying deep learning models directly on edge devices is trending thanks to the advancement of both effective and light-weight deep learning models and energy-saving deep learning processors and accelerators~\cite{chen2014diannao}.

\begin{figure}[ht]
\centering{
        \includegraphics[width=1.0\linewidth]{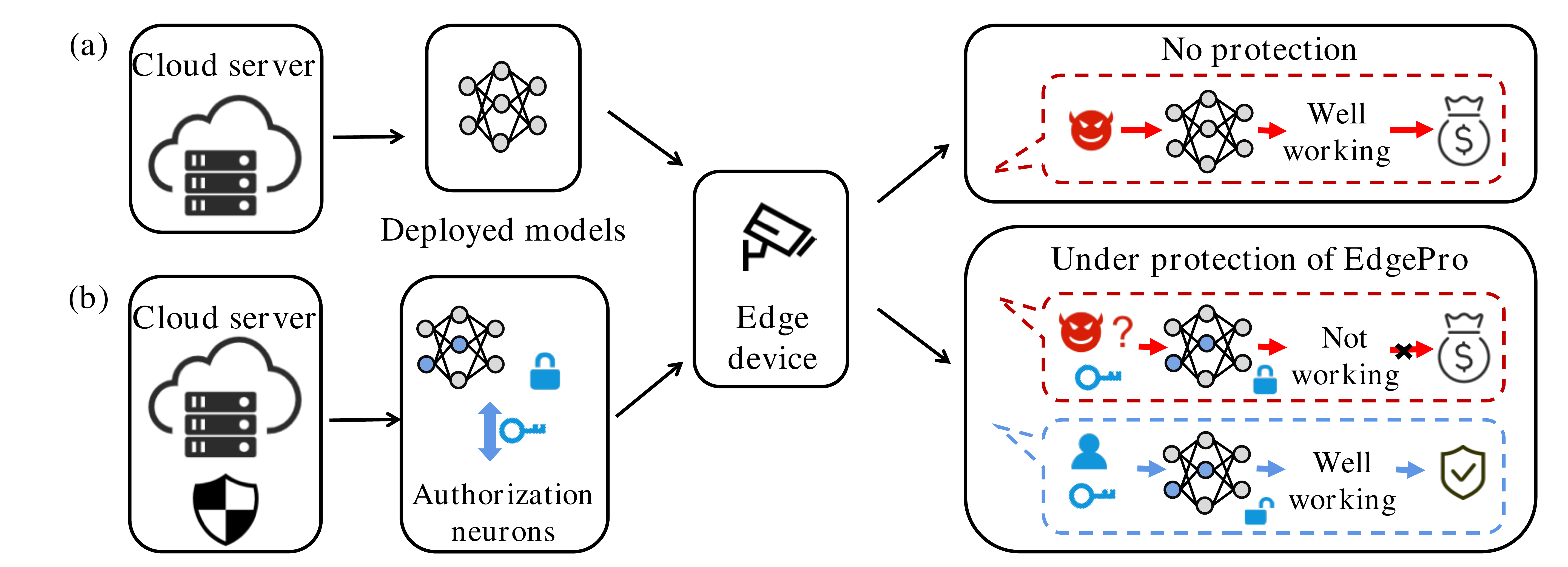} }
\caption{EdgePro embeds the ``passwords'' (i.e., the neuron activation values are locked to the locking values) into the model neurons to ensure that the model cannot work correctly without the ``passwords'', which can protect the edge device model in a light-weight manner.}
\label{fig:ob}
\end{figure}

However, edge device models are vulnerable to illegitimate access by adversaries~\cite{hua2018reverse}.
As shown in Fig.~\ref{fig:ob}(a), when there is no protection for the edge device model, an attacker can easily access the edge device, copy the model and use or sell it for profits.
In order to prevent being stolen or abused, the edge device model need to be carefully protected.

The existing work on edge device model protection can be roughly categorized into hardware-based and software-based protection. The hardware-based protections use trusted execution environments (TEE)~\cite{mo2020darknetz,jang2016privatezone,nakai2021towards,evtyushkin2016flexible} to build a trust region on the main processor, which ensures the models stored in the trust region can run safely. However, TEE is required to swap pages between secure and unprotected memory frequently, which incurs significant overhead.
Other hardware-based protections that set proprietary encryption chips to store the models~\cite{huang2021secure,zuo2020sealing}.
However, they require special customization for different edge devices, which makes them difficult to be widely applied due to the large variety of IoT devices.
Software-based protections use provably secure cryptographic methods to encrypt the models~\cite{xu2021nn,fiore2019multi,rouhani2018deepsecure,ball2019garbled}. The encryption and decryption theoretically ensure the security property.
Unfortunately, the time cost to dynamically decrypt a large number of model parameters on edge devices makes this line of approaches impractical~\cite{trappe2015low,bryzek2013roadmap,guin2018secure}.


There are four challenges in edge device model protection on edge devices, which can also be seen as four requirements.
The protection should
(1) effectively protect models from unauthorized usage;
(2) have little impact on the accuracy of the models;
(3) be light-weight to run smoothly on the resource-limited edge devices;
(4) be robust even if the adversaries know about the existence of the protection, i.e., robust against adaptive attacks.

To overcome the above challenges, one of our intuitions is to add special markers to the input and train a model that can work only when special markers are part of the input.
However, adding markers to input may increase the burden of input preprocessing on the edge devices.
We have observed that the special markers on the input will indirectly cause the changes in neuron activation values during our exploration.
In fact, the objective can also be achieved directly by controlling the changes of neurons.

Therefore, it motivates us to propose a new \underline{\emph{Edge}} device model \emph{\underline{Pro}tection} method, EdgePro, from the perspective of neurons.
The core idea is to leverage the activation values of a small number of neurons as markers, which are called authorization neurons.
If the activation values of these neurons are not met with the locking values during the model inference, it is considered an unauthorized model inference, and vice versa, as shown in Fig.~\ref{fig:ob}(b).

Specifically, we randomly select authorization neurons to guarantee their unpredictability, then lock their activation values as special markers during training, and scaling the activation values of each layer.
Through lock training, we guarantee that neurons in the model have distinct activation states when authorization neurons are locked or not.
Afterward, during the inference, to infer the input using EdgePro trained models, EdgePro only requires to set the activation values of authorization neurons to the locking values.
No other runtime cost is introduced, which is light-weight and practical for running on edge devices.
Moreover, the impact of EdgePro on accuracy is negligible by ensuring the model converges during the training.
The inference time increase of EdgePro is only 60\% of state-of-the-art (SOTA) methods, and the accuracy loss is less than 1\%.
We evaluate the robustness of EdgePro against adaptive attacks, the experimental results of which show strong robustness against reverse engineering, model pruning and model fine-tuning.
Last but not the least, experiments on graph datasets show that EdgePro is general.

To summarize, the contributions of this paper are as follows,

\begin{itemize}
\item As far as we know, from the perspective of neurons, we first time propose a light-weight method, EdgePro, which can protect the edge device models from unauthorized usage from the perspective of neurons.

\item The experimental results demonstrate that EdgePro can protect the models well. The accuracy loss of the EdgePro trained model is only around 1\%, and the inference time increase of EdgePro is only 60\% of other SOTA methods.

\item EdgePro is further evaluated under three adaptive attacks, e.g., reverse engineering, model pruning and model fine-tuning, the results of which demonstrate strong robustness against these adaptive attacks.
\end{itemize}

\section{Related Works}

\subsection{Hardware Protection of Edge Device Models}
An important way to protect the edge device models on edge devices is hardware root-of-trust~\cite{tehranipoor2011introduction,costan2016intel}.
They suggested that a complete edge device model should be implemented in the TEE to protect the confidentiality and integrity of models.
Nakai \textit{et al.}~\cite{nakai2021towards} extended TEE from Intel's SGX to ARM's TrustZone, which is more suitable for edge devices.
Due to the limited storage of TEE, Gangal \textit{et al.}~\cite{gangal2020hybridtee} partitioned an edge device model and encapsulated only some layers in SGX powered TEE.
However, the model parameters stored in unprotected memory are still easy to be stolen, and adversaries can build a complete model through model reverse engineering~\cite{huang2020new,regazzoni2020machine}.
Another idea is to explore building custom secure neural network accelerators~\cite{isakov2018preventing,guin2018secure,li2019p3m}.
They used the Physical Unclonable Function and Processing-In-Memory to ensure that the model can be decrypted only for the authorized devices.

In addition, some work~\cite{chakraborty2020hardware,goldstein2021preventing,hashemi2021darknight} also propose to protect the model parameters by confusing the storage of the model.
Cammarota \textit{et al.}~\cite{chakraborty2020hardware} proposed HANN, a hardware-assisted model protection approach.
They obfuscated the weights of the model based on a secret key that is stored in a trusted hardware device. Users can use the model only if they can provide the trusted key device.
Goldstein \textit{et al.}~\cite{goldstein2021preventing} proposed a solution based on hardware root of trust and public key cryptography infrastructure, which defends against model theft during model distribution and deployment/execution via light-weight, keyed model obfuscation scheme.
Similarly, Hashemi \textit{et al.}~\cite{hashemi2021darknight} provided provable model security by creating input obfuscation in TEE using a custom data encoding strategy based on matrix masks.


\subsection{Software Protection of Edge Device Models}
The software protection of edge device models is realized in two forms: encryption algorithm and intellectual property protection.
In intellectual property protection, Tang \textit{et al.}~\cite{tang2020deep} proposed a serial number-based model protection method, which uses the knowledge distillation to assign a serial number to the customer (student) model, and the customer model can be used normally only if the correct serial number is input.
Chen and Wu \textit{et al.}~\cite{chen2018protect} designed an adversarial example-based transformation module to provide empowered inputs. When an unauthorized user provides input to the model, it is perturbed by adversarial perturbations, resulting in poor performance.
Fan \textit{et al.}~\cite{fan2019rethinking} proposed embedding a specific passport layer in the model, which can paralyze the functionality of the neural network if unauthorized use, or maintain its functionality if verified.
Zhang \textit{et al.}~\cite{zhang2020passport} also proposed a passport-aware normalization paradigm for model protection. A new passport-aware branch was added and trained along with the model. The model performance can be maintained only if the correct passport is provided, otherwise it will drop significantly~\cite{zhang2020passport}.
Alam \textit{et al.}~\cite{alam2020deep} utilized s-boxes with cryptographic properties to lock parameters of a DNN model without causing significant increase in inference time and model scales.
Pyone \textit{et al.}~\cite{alam2020deep} used block pixel shuffling with a key as a preprocessing technique to input images, and the protected model was built by training with such preprocessed images.

On the research of encryption algorithm for edge devices, Lu \textit{et al.}~\cite{lu2018new} developed a secure query scheme with high communication efficiency in the fog environment, to ensure that both cloud and edge devices can use it for privacy protection.
Fiore \textit{et al.}~\cite{fiore2019multi} developed a multi-bond homomorphic authenticator from the perspective of data outsourcing, which is suitable for resource-constrained devices.
In addition, there are some methods to encrypt the model and input on edge devices based on Yao's Garbled Circuits~\cite{rouhani2018deepsecure,ball2019garbled}.
However, the existing encryption algorithms still face the problems of high computational cost, complex decryption processes and low efficiency.
This motivates us to propose a new light-weight model protection method, which can protect the model by binding the inputs and outputs.

\section{Threat Model}
\label{sec:threat}

Our objective is to design a light-weight model protection method that can protect models from unauthorized use in an untrusted environment.
For adversaries, they aim to gain benefits from the stolen models deployed on edge devices. They can fully access the memory or execution environment on edge devices at any time, which may be from the actual user, malicious third-party software installed on the device, or a malicious or infected operating system.
More demanding, we assume the adversaries know the existence of EdgePro.
They may design specific adaptive attacks to crack EdgePro by using state-of-the-art techniques, e.g., model fine-tuning, reverse engineering and model pruning.
Specifically, the adaptive attacks are detailed below.

\begin{itemize}
\item \textbf{Model Fine-tuning Attack}: In practice, for an adversary who lacks training data and intends to break EdgePro, one of the easiest ways is to fine-tune the stolen model\cite{shin2016deep,pittaras2017comparison,yosinski2014transferable}.
Model fine-tuning attack is an intentional attack performed by an adversary who tries to invalidate the authorization neurons, which can be regarded as one of the most threatening attacks on EdgePro.
In general, fine-tuning can produce a new model with less extra training data based on the stolen model.
In this way, the new model can inherit the performance of the stolen model, but also forget the previous training information~\cite{liu2018fine}.

\item \textbf{Reverse Engineering Attack}: The key of EdgePro is authorization neurons, their locking values and scale factors.
Once the authorization neurons, locking values and scale factors are exposed, adversaries can use the EdgePro trained model normally.
The reverse engineering attack is such an threat that may be able to excavate critical information.
An adversary with sufficient knowledge of EdgePro may try to look for several authorization neurons by adjusting the activation values of neurons one by one and observing whether the output of the model changes.

\item \textbf{Model Pruning Attack}: Model pruning~\cite{han2015learning,molchanov2016pruning,sze2017efficient} is a technique to reduce the computational overhead of executing a neural network and still keep the performance of the original model by removing redundant neurons.
An adversary may prune and aim to remove the authorization neurons embedded in the model to invalidate EdgePro. Ideally, the adversary may obtain a model with high classification accuracy, and use the model normally after pruning.

\end{itemize}


\section{Methodology}

\begin{figure*}[t]
\centering{
        \includegraphics[width=1.0\linewidth]{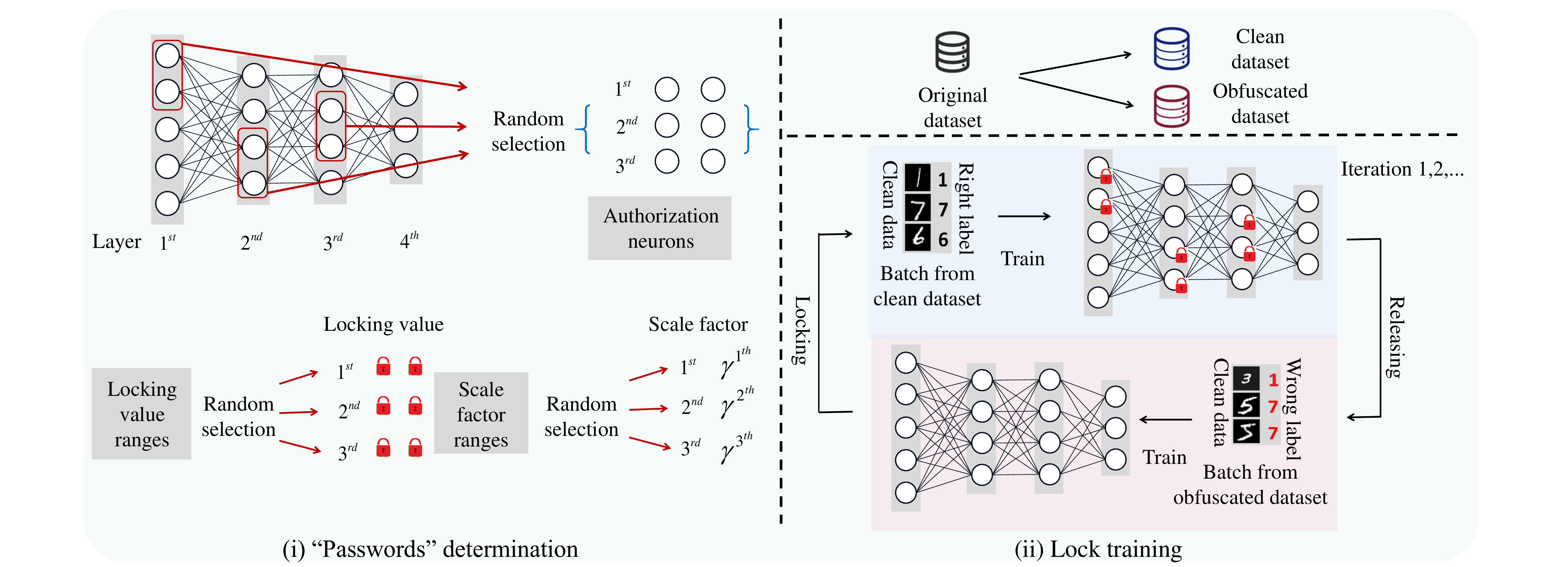} }
\caption{An overview of EdgePro, it includes two parts: ``passwords'' determination and lock training. ``Passwords'' determination is responsible for selecting authorization neurons, locking values and scale factors.
Lock training is responsible for training preparation and lock training for different objectives.
}
\label{fig:ob2}
\end{figure*}



EdgePro consists of two stages, i.e., ``passwords'' determination and lock training.
The overview of EdgePro is presented in Fig.~\ref{fig:ob2}.
The ``passwords'' of EdgePro consist of three parts: authorization neurons, locking values and scale factors. Before locking training, the protector needs to determine these three parts, as shown in Fig.~\ref{fig:ob2} (i).
The second stage is the lock training, as shown in Fig.~\ref{fig:ob2} (ii).
First, an obfuscated dataset is created for lock training.
Then, during the lock training, by alternately locking and releasing authorization neurons, while adjusting different training objectives, EdgePro helps the model adapt to the locking of authorization neurons and eventually converge.
After the model is well trained, if the activation values of authorization neurons do not reach the locking values in the inference phase, the EdgePro trained model will not work properly.

\subsection{``Passwords'' Determination}

EdgePro embeds activate authorization mechanisms, i.e., specific neurons need to meet corresponding activation values, to make sure it will be extremely difficult for unauthorized usage of stolen models.
Therefore, these neurons, which we define as authorization neurons, will be part of the ``passwords'' in the model.
In addition to authorization neurons, the ``passwords'' of the model are also composed of the authorization neuron activation values, which we define as locking values, and the scale factors of neuron activation values in each layer.

In order to ensure unpredictability, we consider that the choice of ``passwords'' should be random, without any strategies.
Therefore, EdgePro will randomly select a small number of neurons in each layer as authorization neurons.
We use $\rho$ to express the proportion of neuron selection authorized by each layer.
In view of the learnability of neural network~\cite{prakash2019repr,han2016dsd}, when $\rho$ is small, the impact on model performance after authorized is limited (we verified the impact of $\rho$ ratio on model performance in Section 5.7.1).
This is the same for the locking values and scale factors, we randomly select them from specific ranges.
As for the specific range, we prove in experiments that an appropriate range will not affect the performance of the model.

Assume $\mathcal{N}$ is the set of all neurons in the $k$-th layer, $\mathcal{A}^k$ is the set of authorization neurons in the $k$-th layer, and $\mathcal{V}^k$ is the set of locking values corresponding to authorization neurons.
When an input example is fed to the model, the $i$-th neuron activation value ${\alpha}^{k+1}_{i}$ in the ($k+1$)-th layer can be expressed as:
\begin{equation}
    \begin{array}{ll}
        {\alpha}^{k+1}_{i} = & \gamma_{k+1} \cdot (Relu((\sum_{j \mid j \in \mathcal{N} \wedge j \notin  \mathcal{A}^k}^{}  w_{j,i}^{k} \cdot {\alpha}^{k}_{j} + \beta^{k}_i) \\
    &+ (\sum_{j \mid j \in \mathcal{N} \wedge j \in  \mathcal{A}^k}^{} v_j^k) )
\end{array}
\label{equ:neurons activation}
\end{equation}
where ${\alpha}^{k}_{j}$ represents the $j$-th neuron activation value in the \textit{k}-th layer, $w_{j,i}$ corresponds to the weight between the $i$-th neuron and the $j$-th neuron, and $\beta^{k}_i$ represents the bias.
Here we use $Relu$ as the activation function for demonstration because of its non-negative property.
$\gamma_{k+1}$ is the scale factor of the ($k+1$)-th layer.
Eq.~(\ref{equ:neurons activation}) means that when ${\alpha}^{k}_{j}$ is not an authorization neuron, its activation value will be calculated by weight. When ${\alpha}^{k}_{j}$ is an authorization neuron, EdgePro will discard its original activation value and replace it with a locking value $v_j^k$, regardless of any input.

\subsection{Lock Training}

\subsubsection{Training Preparation}
During training preparation, EdgePro needs to divide the training dataset.
EdgePro divides the training dataset $D$ into two equal parts, clean dataset $D_c$ and obfuscated dataset $D_o$.
Clean dataset $D_c$ includes clean data $x$ and its right label $y$. Obfuscated dataset $D_o$ includes clean data $x$ and the wrong label $y^{\prime}$ which is modified to a random uncertain label.
Here we guarantee $D_c \cap D_o = \emptyset$ and $D_c \cup D_o = D$.

\subsubsection{Lock Training Process}
In lock training, each iteration contains two batches of training. First, EdgePro samples a batch from the clean dataset $D_c$, and sets the activation values of authorization neurons to the locking values for training.
In this training, the output of each layer will be scaled by scale factors.
Then EdgePro releases the authorization neurons and samples a batch from the obfuscated dataset for training.
EdgePro performs iterative training until the model converges on the clean dataset.

We use the function $R$ to represent the operation of setting the activation values of authorization neurons to the locking values and scaling the output.
Note that we do not do anything with the obfuscated dataset.
EdgePro's training objective can be expressed as:
\begin{equation}
\mathop{\arg\max}(\sum_{x_i \in D_c} P [G^{t} (R(x_{i}) = y_{i})] \notag + \sum_{x_j \in D_o} P [G^{t} (x_{j} \ne y_j)])
\label{equ:dataset train model}
\end{equation}
where $G^t$ can both assign the highest probability to label $y_i$ for data $x_i$ in the $D_c$ and the lowest probability to label $y_j$ for data $x_j$ in the $D_o$.
EdgePro's training objective is for the model to achieve the highest classification accuracy when the activation values of authorization neurons are set to the locking values, and vice versa.
In this way, we embed the active authorization mechanism in the EdgePro trained model.
We only need to encrypt and store the authorization neurons, their locking values, and scale factors, instead of the entire model weights to protect the model.


\section{Experiments}
\subsection{Experimental Setup}
\textbf{Datasets and Models.}

\textbf{(1) MNIST~\cite{lecun1998gradient}\footnote{MNIST can be downloaded at \emph{http://yann.lecun.com/exdb/mnist/}}} is a general image classification dataset which contains 70,000 handwritten gray-scale digital images with size of 28x28, ranging from 0 to 9 (10 classes).

\textbf{(2) CIFAR-10~\cite{krizhevsky2012imagenet}\footnote{CIFAR-10 can be downloaded at \emph{ https://www.cs.toronto.edu/~kriz/  cifar.html}}} is a general image classification dataset which contains 60,000 RGB color images with size of 32x32 in 10 classes.
Each pixel includes three RGB values, with an integer value in [0, 255].

\textbf{(3) CIFAR-100~\cite{krizhevsky2012imagenet}\footnote{CIFAR-100 can be downloaded at \emph{ https://www.cs.toronto.edu/~kriz/  cifar.html}}} is a general image classification dataset which contains 60,000 RGB color images with size of 32x32 in 100 classes, which contains 50,000 training images and 10,000 testing images.

\textbf{(4) Tiny-ImageNet~\cite{deng2009imagenet}\footnote{Tiny-ImageNet can be downloaded at \emph{ http://cs231n.stanford.edu/tiny-imagenet-200.zip}}} is a computer vision dataset containing 200 classes.
Each class has 500 training examples, 50 testing examples and 50 valid examples.

For each dataset, we use different network architectures for experiments.
On MNIST, we adopt LeNet~\cite{lecun1998gradient} and MLP.
We adopt ResNet~\cite{he2016identity}, VGG~\cite{simonyan2014very} for CIFAR-10, and ResNet~\cite{he2016identity}, VGG~\cite{simonyan2014very}, DenseNet~\cite{huang2017densely} for CIAFR-100.
On Tiny-ImageNet, experiments are implemented on
DenseNet~\cite{huang2017densely}, SENet~\cite{hu2018squeeze} and ShuffleNet~\cite{zhang2018shufflenet}.
Model configurations and experiment parameter setups are summarized in Table~\ref{tab:dataset and model}, which records the learning rate, batch size, and training run epoch used for each dataset.

\begin{table}[ht]
\centering
\caption{The model configurations and experiment parameter setups.}
\label{tab:dataset and model}
\resizebox{\linewidth}{!}{%
\begin{tabular}{ccccc}
\hline
\textbf{Datasets}                        & \textbf{Models}        & \textbf{Learning rate}          & \textbf{Batch size}           & \textbf{Epoch}               \\ \hline \hline
\multirow{3}{*}{MNIST}         & LeNet-1      & \multirow{3}{*}{0.01}  & \multirow{3}{*}{64}  & \multirow{3}{*}{20}  \\
                               & LeNet-5      &                        &                      &                      \\
                               & MLP          &                        &                      &                      \\ \hline
\multirow{3}{*}{CIFAR-10}      & ResNet-18    & \multirow{3}{*}{0.01}  & \multirow{3}{*}{64}  & \multirow{3}{*}{40}  \\
                               & ResNet-50    &                        &                      &                      \\
                               & VGG-16       &                        &                      &                      \\ \hline
\multirow{3}{*}{CIFAR-100}     & ResNet-101       & \multirow{3}{*}{0.001} & \multirow{3}{*}{128} & \multirow{3}{*}{100} \\
                               &  VGG-19  &                        &                      &                      \\
                               & DenseNet-121 &                        &                      &                      \\ \hline
\multirow{3}{*}{Tiny-ImageNet} & DenseNet-121 & \multirow{3}{*}{0.001} & \multirow{3}{*}{128} & \multirow{3}{*}{200} \\
                               & SeNet        &                        &                      &                      \\
                               & ShuffleNet   &                        &                      &                      \\ \hline
\end{tabular}%
}
\end{table}

\begin{table*}[t]
\centering
\caption{Evaluation results of EdgePro on different datasets and model architectures, including the test accuracy for the normally trained model, and for EdgePro trained models. For EdgePro, the values without brackets in the table represent $acc_{nl}$, and the values with brackets represent $acc_{nu}$.}
\label{tab:Experiment 1}
\resizebox{\textwidth}{!}{%
\begin{tabular}{ccc|ccc|ccc|ccc}
\hline
\multicolumn{3}{c|}{\textbf{MNIST}}                                                           & \multicolumn{3}{c|}{\textbf{CIFAR-10}}                                                          & \multicolumn{3}{c|}{\textbf{CIFAR-100}}                                                           & \multicolumn{3}{c}{\textbf{Tiny-ImageNet}}                                                        \\ \hline
\multicolumn{1}{c|}{Models}                    & Normal                 & EdgePro   & \multicolumn{1}{c|}{Models}                      & Normal                 & EdgePro   & \multicolumn{1}{c|}{Models}                         & Normal                 & EdgePro  & \multicolumn{1}{c|}{Models}                         & Normal                 & EdgePro  \\ \hline \hline
\multicolumn{1}{c|}{\multirow{2}{*}{LeNet-1}} & \multirow{2}{*}{98.48\%} & 97.60\%   & \multicolumn{1}{c|}{\multirow{2}{*}{ResNet-18}} & \multirow{2}{*}{87.28\%} & 87.10\%   & \multicolumn{1}{c|}{\multirow{2}{*}{VGG-19}}       & \multirow{2}{*}{73.77\%} & 70.52\%  & \multicolumn{1}{c|}{\multirow{2}{*}{DenseNet-121}} & \multirow{2}{*}{55.47\%} & 55.10\%  \\
\multicolumn{1}{c|}{}                         &                          & (17.67\%) & \multicolumn{1}{c|}{}                           &                          & (11.71\%) & \multicolumn{1}{c|}{}                              &                          & (1.00\%) & \multicolumn{1}{c|}{}                              &                          & (0.35\%) \\ \hline
\multicolumn{1}{c|}{\multirow{2}{*}{LeNet-5}} & \multirow{2}{*}{100.00\%}   & 99.29\%   & \multicolumn{1}{c|}{\multirow{2}{*}{ResNet-50}} & \multirow{2}{*}{89.05\%} & 88.72\%   & \multicolumn{1}{c|}{\multirow{2}{*}{ResNet-101}}   & \multirow{2}{*}{75.98\%} & 75.11\%  & \multicolumn{1}{c|}{\multirow{2}{*}{SENet}}        & \multirow{2}{*}{56.80\%} & 55.82\%  \\
\multicolumn{1}{c|}{}                         &                          & (10.01\%) & \multicolumn{1}{c|}{}                           &                          & (7.05\%)  & \multicolumn{1}{c|}{}                              &                          & (0.68\%) & \multicolumn{1}{c|}{}                              &                          & (0.61\%) \\ \hline
\multicolumn{1}{c|}{\multirow{2}{*}{MLP}}     & \multirow{2}{*}{98.88\%} & 98.79\%   & \multicolumn{1}{c|}{\multirow{2}{*}{VGG-16}}    & \multirow{2}{*}{89.10\%} & 88.92\%   & \multicolumn{1}{c|}{\multirow{2}{*}{DenseNet-121}} & \multirow{2}{*}{74.71\%} & 74.20\%  & \multicolumn{1}{c|}{\multirow{2}{*}{ShuffleNet}}   & \multirow{2}{*}{56.20\%} & 55.45\%  \\
\multicolumn{1}{c|}{}                         &                          & (12.16\%) & \multicolumn{1}{c|}{}                           &                          & (13.15\%) & \multicolumn{1}{c|}{}                              &                          & (1.03\%) & \multicolumn{1}{c|}{}                              &                          & (0.40\%) \\ \hline
\end{tabular}%
}
\end{table*}


\textbf{Evaluation Metrics.}
The metrics used in the experiments are defined as follows:
\begin{itemize}
\item Neuron locking accuracy ($acc_{nl}$): $acc_{nl} = \frac{n_{nl}}{N} $, where $n_{nl}$ is the number of examples correctly classified by the model when it is authorized, $N$ is the total number of examples.

\item Neuron unlocking accuracy ($acc_{nu}$): $acc_{nu} = \frac{n_{nu}}{N} $, where $n_{nu}$ is the number of examples correctly classified by the model when the model is not authorized, $N$ is the total number of examples.

\end{itemize}
The larger the gap between $acc_{nl}$ and $acc_{nu}$ means the better the protective effect of EdgePro.

\textbf{Baselines.}
We implement and compare four SOTA methods with EdgePro to evaluate their performance, including Password Normalization (PN)~\cite{zhang2020passport},
Deep-Lock~\cite{alam2020deep}, AntiP~\cite{chen2018protect}, and LIE~\cite{sirichotedumrong2019privacy}.
All baselines are advanced protection methods for authorized use of models, and they are configured according to the performance setting reported in the respective papers.

\textbf{Platform.} We leverage a platform with the following setup: CPU is Intel XEON 6240 2.6GHz x 18C, GPU is Tesla V100 32GiB, the Memory is DDR4-RECC 2666 16GiB, the operating system is Ubuntu 16.04, the programming language is Python 3.6.0, and the deep learning framework is PyTorch-1.4.0.




\begin{table}[]
\centering
\caption{Comparison of EdgePro's time cost with normal training and baselines.}
\label{time cost}
\resizebox{\linewidth}{!}{%
\begin{tabular}{lllrc}
\hline
\textbf{Datasets}                                                                   & \textbf{Models}                                                                    & \textbf{Methods}       & \textbf{Size(bit)}    & \textbf{Time(s)} \\  \hline \hline
\multirow{6}{*}{MNIST}                                                    & \multirow{6}{*}{LeNet-5}                                                 & Normal       & 243K    & 0.50    \\
                                                                          &
                                                                          & PN       & 282K    & 0.69    \\
                                                                          &
                                                                          & Deep-Lock    & 486K    & 0.73                                      \\
                                                                          &                                                                          & AntiP & 252K    & 0.63    \\
                                                                          &                                                                          & LIE          & 243K    & 0.65    \\
                                                                          &                                        & \textbf{EdgePro}      & \textbf{243K}    & \textbf{0.62}     \\ \hline
\multirow{6}{*}{CIFAR-10}                                                 & \multirow{6}{*}{ResNet-18}                                               & Normal       & 42.68M  & 0.76    \\
                                                                          &
                                                                          & PN       & 44.60M    & 2.06    \\
                                                                          &
                                                                          & Deep-Lock    & 85.36M  & 1.80        \\
                                                                          &                                                                          & AntiP & 42.70M  & 1.26    \\
                                                                          &                                                                          & LIE          & 42.68M  & 1.44    \\
                                                                          &                                                                            & \textbf{EdgePro}      & \textbf{42.68M}  & \textbf{0.90}  \\ \hline
\multirow{6}{*}{CIFAR-100}                                                & \multirow{6}{*}{VGG-19}                                                  & Normal       & 174.08M & 0.86    \\
                                                                          &
                                                                          & PN       & 186.47M    & 3.70    \\
                                                                          &
                                                                          &  Deep-Lock    & 348.17M & 3.46  \\
                                                                          &                                                                          & AntiP & 174.10M & \textbf{1.76}    \\
                                                                          &                                                                          & LIE          & 174.08M & 2.28    \\
                                                                          &                                                                          &  \textbf{EdgePro}      & \textbf{174.08M} & 1.83    \\ \hline
\multirow{6}{*}{\begin{tabular}[c]{@{}c@{}}Tiny-\\ ImageNet\end{tabular}} & \multirow{6}{*}{\begin{tabular}[c]{@{}c@{}}DenseNet-\\ 121\end{tabular}} & Normal       & 27.78M  & 1.04    \\
                                                                          &
                                                                          & PN       & 29.02M    & 1.90    \\
                                                                          &
                                                                          &  Deep-Lock    & 55.55M  & 1.76    \\
                                                                          &                                                                          & AntiP & 27.82M  & 1.40    \\
                                                                          &                                                                          & LIE          & 27.78M  & 1.66    \\
                                                                          &                                                                          &  \textbf{EdgePro}      & \textbf{27.78M}  & \textbf{1.39}  \\ \hline
\end{tabular}%
}
\end{table}

\subsection{Effectiveness of EdgePro}
\label{Effectiveness of EdgePro}
In this section, we focus on the evaluation results of EdgePro when model is authorized and not authorized.

\textbf{Implementation Details.}
(1) We evaluate EdgePro in two scenarios, including normal training and EdgePro training.
In normal training, we train the model normally then test the model accuracy.
In EdgePro traing, we set $\rho = 5$, which means 5\% neurons in each layer are selected as the authorization neurons, and the each locking value $v$ will be randomly selected in the range $\mathcal{V}=$ (0, 1).
The scale factors $\gamma$ also will be randomly selected in the range (0.2, 1).
(2) We train the model until the specified training epoch is reached, or the model loss is below 1e-4.
To mitigate non-determinism, we repeated the experiment for 5 times and reported the average results.

\textbf{Results and Analysis.}
The results of which are shown in Table~\ref{tab:Experiment 1}.
We can see that EdgePro trained models achieve high $acc_{nl}$ and low $acc_{nu}$.
The low $acc_{nu}$ values indicate that when the activation values of the authorization neurons in the EdgePro trained model do not meet the locking values, the model accuracy is reduced to the level of random guessing.
This reflects that EdgePro discourages illegal users from using models by reducing model performance.
Regarding the $acc_{nl}$, compared with the original accuracy of the normal trained model,
the EdgePro trained model has a minor accuracy loss of 1.63\% on average.
The accuracy loss is understandable and maybe inevitable because EdgePro locks some of the neurons in the model.
Overall, EdgePro has protective effects on the four datasets and for the twelve models.

\subsection{Complexity Comparison of EdgePro}
\label{Time Complexity}
In this section, we focus on how much extra time cost is introduced by EdgePro in inferring phases.

\textbf{Implementation Details.}
(1) We consider that training is offline and usually a one-time process. Therefore, the server has sufficient training time, and we concern about the time cost of the model in the inferring phase.
We do not compare with the hardware method because hardware encryption is compatible with software encryption and EdgePro is basically a software-based protection as well.
(2) EdgePro selected 5\% of the neurons in the model as authorization neurons, then we measure the time cost for the model to infer 1000 examples records.
Meanwhile, we measure the time cost of the baselines for comparison.
For encryption protection method, Deep-Lock, the decryption also needs to be counted in.

\textbf{Results and Analysis.}
The results can be found in Table~\ref{time cost}.
In this experiment, we can see that for the inference phase time cost, EdgePro only takes 36.5\% more than the normal process, while baselines take 164.2\% more time.
This is because the EdgePro authorization only needs to activate a small number of neurons to reach the locking values, which reduces the time overhead.
This means that EdgePro introduces minimal overhead in the inference phase compared to baselines, which can improve model performance and user experience.
Additionally, we observe that the EdgePro on large models add more time overhead.
To this this problem, we consider that for large models, e.g., VGG-19, EdgePro can reduce the extra overhead by using a small $\rho$, e.g., using $\rho = 1$.
Last but not the least, it cannot be ignored that EdgePro does not increase the parameter size of the model, i.e., EdgePro does not introduce additional storage space.
This facilitates the deployment of EdgePro on resource-constrained edge devices.


\subsection{Robustness Comparison of EdgePro}
\label{sec:robustness}
The attacker may try to launch adaptive attacks against model protection methods if they know the existence of model protection methods.
In this section, we compare the robustness between EdgePro and baselines against model fine-tuning attack.
Going a step further, we evaluate the robustness of EdgePro under two adaptive attacks designed specifically for EdgePro (as analyzed in our threat model in Section~\ref{sec:threat}).

\subsubsection{Model Fine-tuning Attack}
Fine-tuning could be an intuitive way for an adversary to remove authorization neurons in an EdgePro trained model with a small amount of data.
Therefore we compare the protection effect of EdgePro with baselines under fine-tuning attack.

\textbf{Implementation Details.}
(1) We reserve 10\% test data for fine-tuning EdgePro trained models.
In particular, the fine-tuning epoch is 10 for the MNIST and CIFAR-10 datasets, and 20 for the CIFAR-100 and Tiny-ImageNet datasets.
(2) We compare the robustness of baselines: PN~\cite{zhang2020passport},
LIE~\cite{sirichotedumrong2019privacy}, and AntiP~\cite{chen2018protect}.
We do not consider Deep-Lock~\cite{alam2020deep} because it cannot be fine-tuning after encryption.
(3) In addition, we consider a new scenario: the scale factors are leaked, i.e. the adversary knows scale factors $\gamma$ used by each layer of EdgePro. We define this scenario as ``EdgePro-$\gamma$'' to measure the robustness of EdgePro.
(4) We use $acc_{nu}$ as a measure of model robustness.
The lower the $acc_{nu}$ of the fine-tuned model, the more robust the method is.


\begin{figure}[ht]
\centering
    \subfigure[MNIST \& LeNet-5]{
        \includegraphics[width=0.45\linewidth]{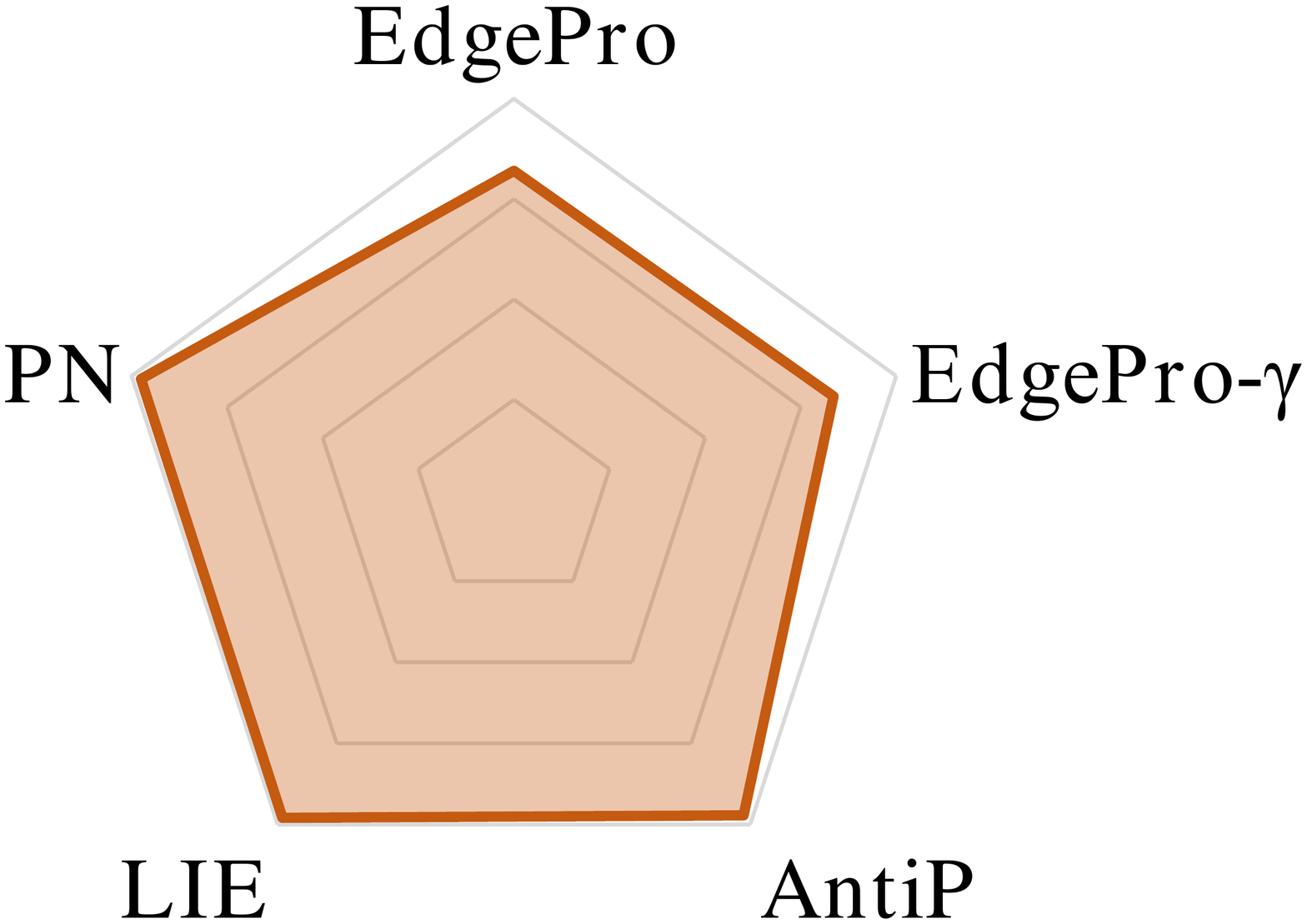} }
    \subfigure[CIFAR-10 \& ResNet-18]{
        \includegraphics[width=0.45\linewidth]{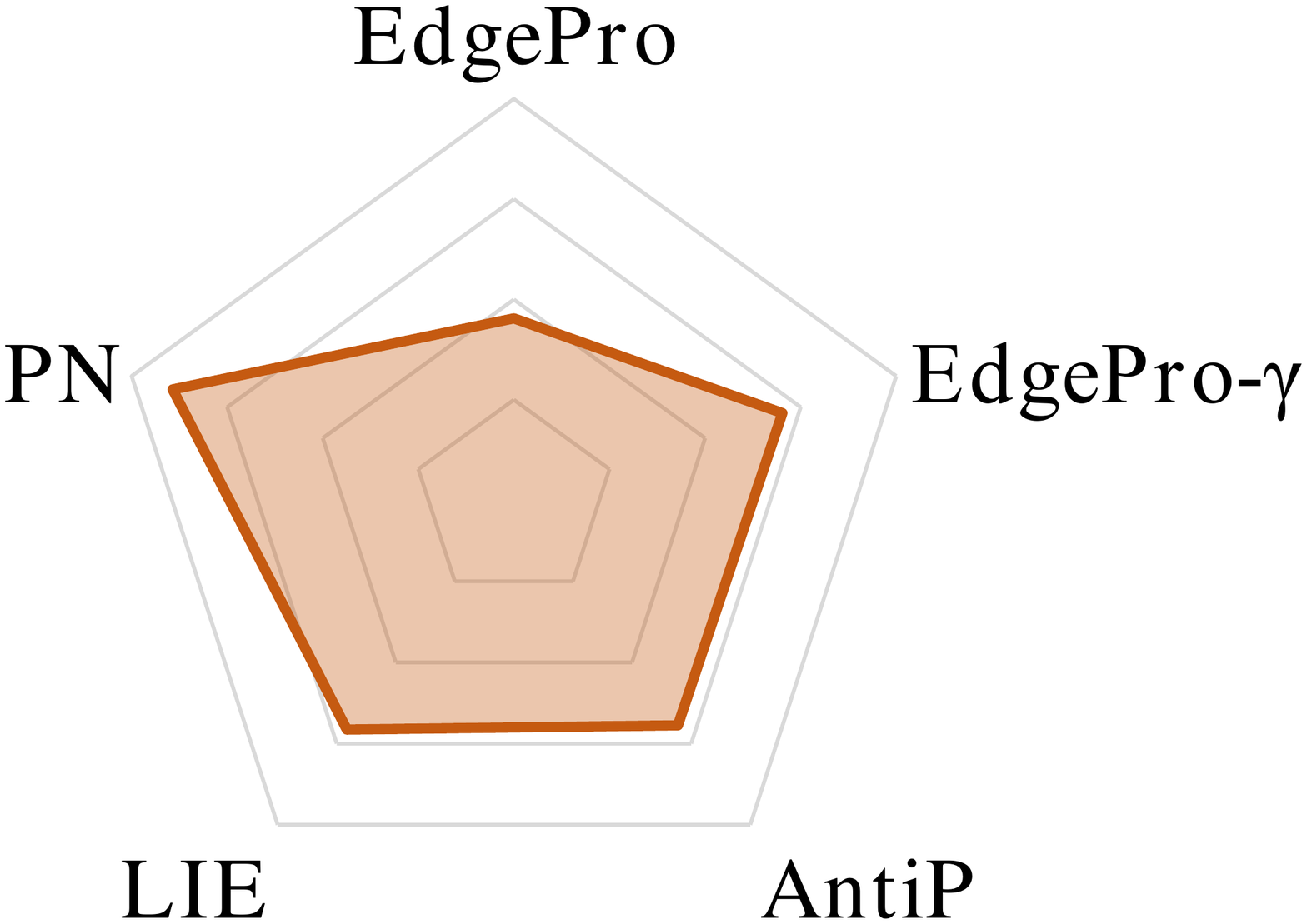} }
    \subfigure[CIFAR-100 \& VGG-19]{
        \includegraphics[width=0.45\linewidth]{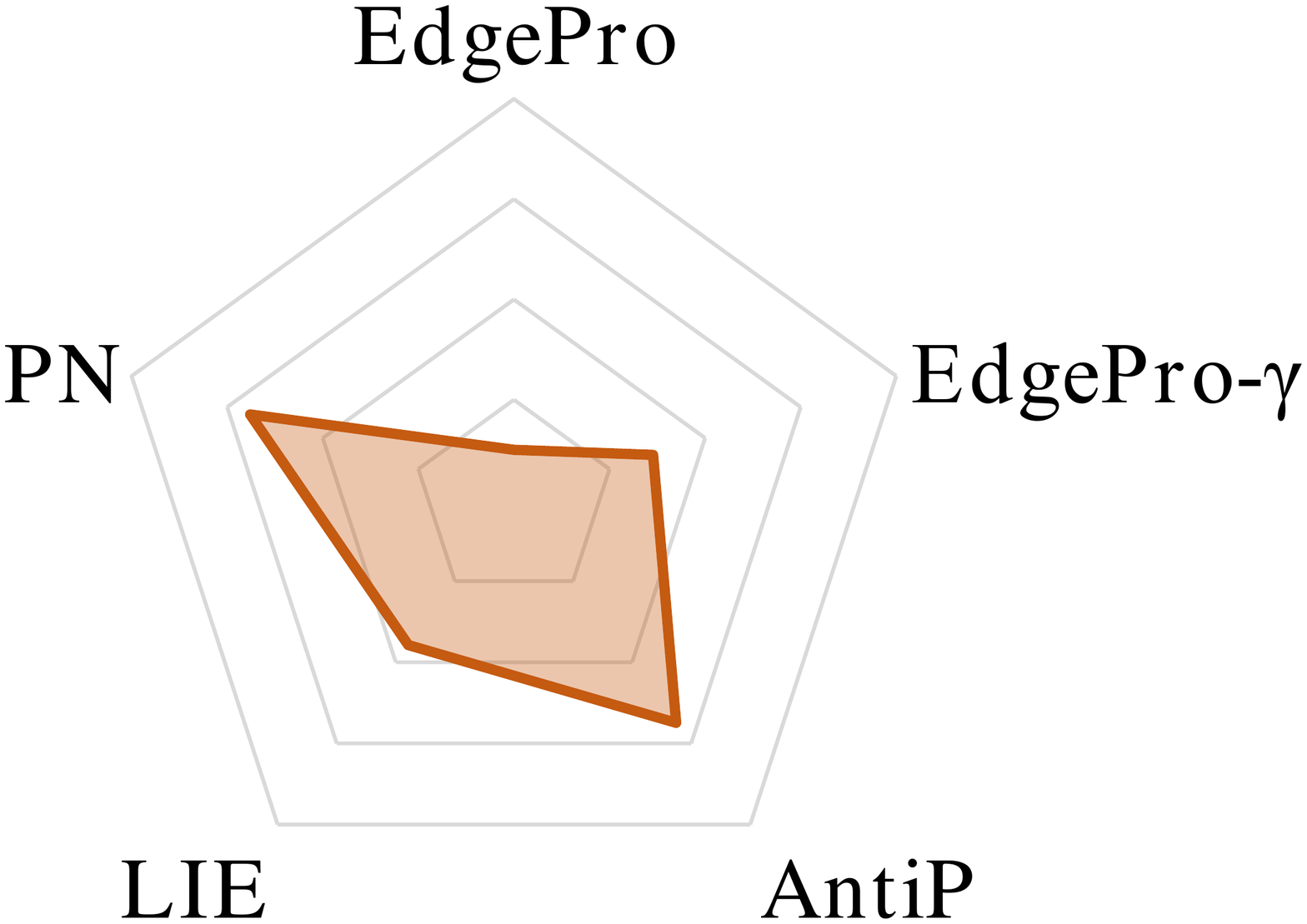} }
    \subfigure[Tint-ImageNet \& DenseNet-121]{
        \includegraphics[width=0.45\linewidth]{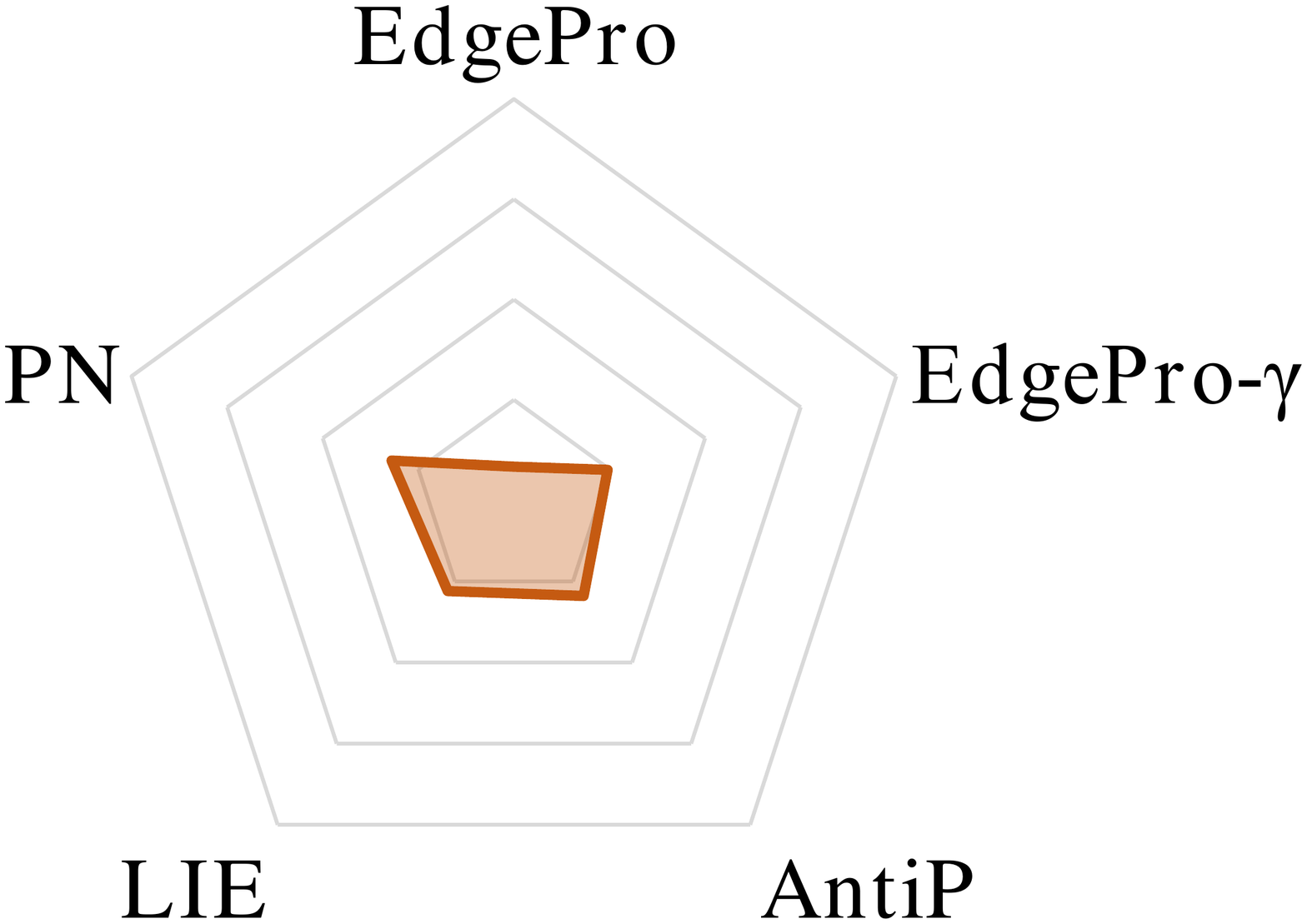} }
    \\
\caption{Robustness evaluation against the fine-tuning attack.}
\label{The rule selecting}
\end{figure}

\textbf{Results and Analysis.}
The experimental results can be observed using the radar chart as shown in Fig~\ref{The rule selecting}.
Compared to baselines, EdgePro is the most robust on the four datasets.
On complex datasets, EdgePro is more robust, e.g., on the CIFAR-100 and Tiny-ImageNet datasets, EdgePro's $acc_{nu}$ is only about 25\% of the baselines.
Unlike baselines that aim to perturb the input or perturb the hidden layer output, EdgePro introduces scale factors into the model.
In this way, EdgePro not only affects the classification layer of the model, but also controls the feature extraction effect of the model.
When the model is not authorized, the scale factors corrupt the output of the model in each layer, thereby enhancing the robustness of EdgePro.
Additionally, comparing different datasets, we find that the models (VGG-19 and DenseNet-121) on large datasets (CIFAR-100 and Tiny-ImageNet) are more robust than the models (LeNet-5 and ResNte-18) on small datasets (MNIST and CIFAR-10).
We speculate the reason is that training on large datasets is more difficult than on small datasets, so the impact of fine-tuning is limited under the condition of a certain amount of data (10\% of the test data).
For ``EdgePro-$\gamma$'', it can be seen that when the attacker masters the scale factors of the model, the robustness of ``EdgePro-$\gamma$'' is less robust than EdgePro, but still better than baselines.
``EdgePro-$\gamma$'' is still able to control model accuracy to an unusable level and has the effect of preventing illegal use of the model.

\begin{table*}[ht]
\centering
\caption{Evaluating the robustness of EdgePro under reverse engineering attack by $acc_{nu}$, $acc_{nl}$ and time.}
\label{tab:reverse engineering}
\resizebox{0.9\textwidth}{!}{%
\begin{tabular}{cccccclr}
\hline
\textbf{Datasets \& Models}                            & \textbf{AZ neurons}                                     & \textbf{$acc_{nu}$}                           & \textbf{$acc_{nl}$}                           & \textbf{Knowledge} & \textbf{RE neurons} & \textbf{$acc_{nu}$} & \textbf{Time(s)} \\ \hline \hline
\multicolumn{1}{c|}{\multirow{6}{*}{MNIST \& LeNet-5}} & \multicolumn{1}{c|}{\multirow{2}{*}{4:29:0.7}}          & \multicolumn{1}{c|}{\multirow{2}{*}{10.70\%}} & \multicolumn{1}{c|}{\multirow{2}{*}{99.18\%}} & All                & 4:1:0.5             & 97.90\%             & 16.32            \\ \cline{5-8}
\multicolumn{1}{c|}{}                                  & \multicolumn{1}{c|}{}                                   & \multicolumn{1}{c|}{}                         & \multicolumn{1}{c|}{}                         & Half               & 4:2:0.6             & 98.10\%             & 28.50            \\ \cline{2-8}
\multicolumn{1}{c|}{}                                  & \multicolumn{1}{c|}{\multirow{2}{*}{5:76:0.2}}          & \multicolumn{1}{c|}{\multirow{2}{*}{10.42\%}} & \multicolumn{1}{c|}{\multirow{2}{*}{99.25\%}} & All                & 5:76:0.1            & 98.63\%             & 1,053.64         \\ \cline{5-8}
\multicolumn{1}{c|}{}                                  & \multicolumn{1}{c|}{}                                   & \multicolumn{1}{c|}{}                         & \multicolumn{1}{c|}{}                         & Half               & 5:76:0.1            & 98.39\%             & 1,666.22         \\ \cline{2-8}
\multicolumn{1}{c|}{}                                  & \multicolumn{1}{c|}{\multirow{2}{*}{4:29:0.7+5:76:0.2}} & \multicolumn{1}{c|}{\multirow{2}{*}{10.20\%}} & \multicolumn{1}{c|}{\multirow{2}{*}{99.48\%}} & All                & 4:29:0.7+5:76:0.7   & 96.25\%             & 48,065.76        \\ \cline{5-8}
\multicolumn{1}{c|}{}                                  & \multicolumn{1}{c|}{}                                   & \multicolumn{1}{c|}{}                         & \multicolumn{1}{c|}{}                         & Half               & Timeout             & Timeout             & Timeout          \\ \hline
\end{tabular}%
}
\end{table*}

\begin{table*}[ht]
\centering
\caption{Evaluating the robustness of EdgePro under model pruning attack by $acc_{nu}$, where $acc$-P20\% represents the $acc$ after pruning 20\% nuerons.}
\label{tab:pruning tabel}
\resizebox{0.75\textwidth}{!}{%
\begin{tabular}{c|c|cccc}
\hline
{\color[HTML]{000000} \textbf{Datasets \& Models}} & {\color[HTML]{000000} \textbf{Metrics}} & {\color[HTML]{000000} \textbf{$acc_{nu}$}} & {\color[HTML]{000000} \textbf{$acc_{nl}$}} & {\color[HTML]{000000} \textbf{$acc_{nu}$-P20\%}} & {\color[HTML]{000000} \textbf{$acc_{nu}$-P60\%}} \\ \hline \hline
                                         & AvgAct                    &                            &                            & 23.50\%                          & 62.33\%                          \\
                                         & GradCAM                       &                            &                            & 10.53\%                          & 46.27\%                          \\
\multirow{-3}{*}{MNIST \& LeNet-5}         & LRP                           & \multirow{-3}{*}{11.10\%}  & \multirow{-3}{*}{97.90\%}  & 15.66\%                          & 56.05\%                          \\ \hline
                                         & AvgAct                    &                            &                            & 22.03\%                          & 22.53\%                          \\
                                         & GradCAM                       &                            &                            & 21.60\%                          & 20.00\%                          \\
\multirow{-3}{*}{CIFAR-10 \& ResNet-18}    & LRP                           & \multirow{-3}{*}{9.50\%}   & \multirow{-3}{*}{87.61\%}  & 14.11\%                          & 18.17\%                          \\ \hline
                                         & AvgAct                    &                            &                            & 4.97\%                           & 2.63\%                           \\
                                         & GradCAM                       &                            &                            & 1.36\%                           & 1.47\%                           \\
\multirow{-3}{*}{CIFAR-100 \& VGG-19}      & LRP                           & \multirow{-3}{*}{1.00\%}   & \multirow{-3}{*}{73.77\%}  & 6.73\%                           & 5.98\%                           \\ \hline
                                         & Activation                    &                            &                            & 6.45\%                           & 2.50\%                           \\
                                         & GradCAM                       &                            &                            & 3.62\%                           & 3.80\%                           \\
\multirow{-3}{*}{Tiny-ImageNet \& DenseNet-121}          & LRP                           & \multirow{-3}{*}{0.35\%}   & \multirow{-3}{*}{55.10\%}  & 8.00\%                           & 6.80\%                           \\ \hline
\end{tabular}%
}
\end{table*}

\subsubsection{Reverse Engineering Attack}
We evaluate the robustness of EdgePro under reverse engineering attack which is an attack method specifically designed for EdgePro.

\textbf{Implementation Details.}
(1) Considering the high complexity of reverse engineering attack, we experiment on the EdgePro trained model using the MNIST dataset. In the fourth and fifth layers of the LeNet-5 model, which have 120 and 84 neurons, respectively, we select one or two neurons as authorization neurons.
(2) As shown in the Table~\ref{tab:reverse engineering}, ``AZ neurons'' means the ``passwords'' and ``4:29:0.7'' refers to that EdgePro selects the $29$-th neuron in the $4$-th layer as the authorization neuron, and the locking value size is $v = 0.7$.
``RE neurons'' refers to the adversary finding the neurons that can crack the EdgePro.
We set a running time for the reverse engineering attack. When the time exceeds 6,000, the attack will terminate with a ``Timeout''.
(3) We set up two scenarios for the adversary based on the adversary's knowledge. ``All'' means the adversary not only knows the EdgePro, but also knows which layer the authorization neurons are in.
``Half'' means the adversary knows the EdgePro method, but does not know the location of the authorization neurons.
In addition, we also assume that in both scenarios, the adversary knows the scale factors $\gamma$ of EdgePro.

\textbf{Results and Analysis.}
Table~\ref{tab:reverse engineering} shows the attack time cost in each scenario.
When the number of authorization neurons is one, that is approximately 0.5\% of the total neurons, the adversary may crack EdgePro.
The adversary just needs to find the neuron that has a similar effect as the authorization neuron rather than infer the exact authorization neuron, e.g. in Table~\ref{tab:reverse engineering} $acc_{nu}$ reachs 97.90\% when ``passwords'' are ``4:1:0.5''.
When we increase the number of authorization neurons to two, the time cost increases drastically and even exceeds the set time (60,000s).
This reflects that the robustness of EdgePro increases significantly as the number of neurons increases, e.g., when the number of authorized neurons goes from 1 to 2, the average time of reverse engineering attacks increases by 45.6 times.
For a model with a large number of authorization neurons, cracking EdgePro costs far more than training a model from scratch.
In addition, in the real scene, the adversary also needs to reverse the scale factor $\gamma$ of each layer, which will further increase the cost of cracking EdgePro.
Compare with both scenarios, ``All'' and ``Half'', EdgePro always protects the model.

\subsubsection{Model Pruning Attack}
We evaluate the robustness of EdgePro under model pruning attack which is specially designed for EdgePro.

\textbf{Implementation Details.}
(1) We consider three basis metrics for model pruning: Average Activation (AvgAct), GradCAM~\cite{selvaraju2017grad}, and Layer-wise Relevance Propagation (LRP)~\cite{bach2015pixel}. Those metrics can measure the importance of neurons.
(2) We use 10\% of the test data to stimulate the EdgePro-trained model (where $\rho$=5, $\mathcal{V}$=(0, 1)) and record the three metrics of each neuron in the model and and order by metrics.
Then we iteratively prune the neurons of the model according to the ascending order.
Finally, We feed the remaining data to the model to compute $acc_{nu}$.
(3) In Table~\ref{tab:pruning tabel}, we adapt two pruning rates, 20$\%$ and 60$\%$, where ``$acc_{nu}$-P20\%'' represents the $acc_{nu}$ after 20\% of the neurons are pruned.
(4) We use $acc_{nu}$ as a measure of robustness.
The lower the $acc_{nu}$ of the pruned model, the more robust the EdgePro is.

\textbf{Results and Analysis.}
Table~\ref{tab:pruning tabel} shows the results of EdgePro on four datasets.
Under both pruning rates, the $acc_{nu}$-P are all very low, which indicates that the pruning can not crack EdgePro.
For instance, when pruning rate is 60\%, $acc_{nu}$-P60\% is 2.1 times higher than $acc_{nu}$ on average and 6.9 times higher than $acc_{nu}$ on complex datasets, but is still far below the model accuracy of normal use.
What cannot be ignored is when the adversary uses a higher pruning rate, it decreases $acc_{nu}$-P at the same time, making the model unusable.
This shows that the authorization neurons in the model are concealed.
The adversary cannot detect authorization neurons by stimulating neurons, and also cannot obtain the right to use the model by pruning the authorization neurons.
Similar to the experimental results of model fine-tuning attack, EdgePro shows stronger robustness on complex datasets and large models. This is because large models have more authorization neurons.
Comparing the three pruning metrics, none of them can break EdgePro.
This reflects our selection of authorization neurons is sufficient and sufficient, EdgePro is resistant to model pruning attacks.

\subsection{EdgePro Case Study on Graph Dataset}

\begin{table}[]
\centering
\caption{Evaluating the EdgePro on two graph datasets: Cora and PubMed. The values without and with brackets in the table represent $acc_{nl}$ and $acc_{nu}$.}
\label{tab: case study}
\resizebox{\linewidth}{!}{%
\begin{tabular}{ccc|ccc}
\hline
\multicolumn{3}{c|}{\textbf{Cora}}                                                      & \multicolumn{3}{c}{\textbf{PubMed}}                                                     \\ \hline
\multicolumn{1}{c|}{Model}                & Normal                   & EdgePro & \multicolumn{1}{c|}{Model}                & Normal                   & EdgePro \\ \hline \hline
\multicolumn{1}{c|}{\multirow{2}{*}{GCN}} & \multirow{2}{*}{83.40\%} & 83.00\% & \multicolumn{1}{c|}{\multirow{2}{*}{GCN}} & \multirow{2}{*}{84.20\%} & 84.00\% \\
\multicolumn{1}{c|}{}                     &                          & (14.40\%) & \multicolumn{1}{c|}{}                     &                          & (29.80\%) \\ \hline
\multicolumn{1}{c|}{\multirow{2}{*}{SGC}} & \multirow{2}{*}{80.60\%} & 80.00\% & \multicolumn{1}{c|}{\multirow{2}{*}{SGC}} & \multirow{2}{*}{83.60\%} & 83.20\% \\
\multicolumn{1}{c|}{}                     &                          & (15.00\%) & \multicolumn{1}{c|}{}                     &                          & (34.60\%) \\ \hline
\multicolumn{1}{c|}{\multirow{2}{*}{GAT}} & \multirow{2}{*}{89.00\%} & 89.00\% & \multicolumn{1}{c|}{\multirow{2}{*}{GAT}} & \multirow{2}{*}{83.20\%} & 80.40\% \\
\multicolumn{1}{c|}{}                     &                          & (14.80\%) & \multicolumn{1}{c|}{}                     &                          & (32.40\%) \\ \hline
\end{tabular}%
}
\end{table}

The deep learning models deployed on edge devices are diverse, and there are not only for image tasks.
In this section, we discuss the generality of EdgePro, e.g., whether EdgePro has a protection effect on node classification tasks.
Since graph-level anomaly detection has been a promising means in many different domains~\cite{wu2021graph,wu2021forest}, such as transportation, energy, and factory, it is necessary to protect graph neural network (GNN)~\cite{zhang2022semantics,wu2019simplifying} on edge devices.

\textbf{Implementation Details.}
Specifically, we select a small graph dataset Cora and a large graph dataset PubMed.

{(1) Cora~\cite{yang2016revisiting}\footnote{Cora can be downloaded at \emph{http://www.cs.umd.edu/~sen/lbc-proj/LBC.html}}} consists of 2,708 scientific publications classified into one of seven classes. The citation network consists of 5,429 links. Each publication in the dataset is described by a 0/1-valued word vector indicating the absence/presence of the corresponding word from the dictionary. The dictionary consists of 1,433 unique words.

{(2) PubMed~\cite{namata2012query}\footnote{PubMed can be downloaded at
\emph{https://linqs-data.soe.ucsc.edu/public/  Pubmed-Diabetes.tgz}}} consists of 19,717 scientific publications from PubMed database pertaining to diabetes classified into one of three classes. The citation network consists of 44,338 links. Each publication in the dataset is described by a weighted word vector from a dictionary which consists of 500 unique words.

For each dataset, we use three models, GCN~\cite{kipf2016semi}, SGC~\cite{wu2019simplifying} and GAT~\cite{velickovic2017graph} to train.
The learning rates are all 0.01, and a total of 100 epochs are trained. For GCN model, hyperparameters are chosen as
follows: $0.5$ (dropout rate of first and last layer), $5\times 10^{-4}$ (L2 regularization at first layer) and $128$ (number of units for each hidden layer). The SGC model and the GCN model keep the same hyperparameter settings. For a two-layer GAT model, the first layer consists of 8 attention heads computing 8 features each. The second layer is used for classification: a single attention head that computes 128 features, followed by ReLU activation.

\textbf{Results and Analysis.}
Tabel~\ref{tab: case study} shows the results on two datasets.
As we expected, EdgePro also has a protection effect on graph datasets, and the accuracy of the model when unauthorized tends to be close to random guessing.
For example, on the Cora dataset of GAT model, the $acc_{nl}$ is 5.7 times larger than $acc_{nu}$.
This shows that EdgePro is general and competent for different task scenarios.

\subsection{Effect of Authorization Neuron Selection on EdgePro}
EdgePro guarantees unpredictability by randomly selecting authorization neurons.
In this section, we discuss the effect of different selection strategies rather than random selection on EdgePro.

\textbf{Implementation Details.}
(1) According to the works~\cite{zeng2021rethinking,wang2019neural} we use several indicators to describe the importance of a neuron, including its activation value, activation frequency, weight value, Grad-CAM~\cite{selvaraju2017grad} and LRP~\cite{bach2015pixel}.
(2) We use a small number of examples to stimulate the neurons in the pre-trained model, and then rank the neurons according to different indicators.
Our intuition is that the change of low importance neurons has less impact on the performance of the model, and is more suitable to be selected as authorization neurons.
Therefore, after ranking the neurons, we give these neurons corresponding weights, and make the random selection according to the weights.
The neurons with lower importance will be given greater weight, which means that the probability of being selected is higher.
(3) We take \textbf{r}andom \textbf{n}euron \textbf{r}anking (\textbf{RNR}) selection as the baseline, and design five neuron ranking selection strategies.
The training parameters for each strategy will as same as in Section~\ref{Effectiveness of EdgePro}.
The details of five strategies are as follows:

\begin{itemize}

\item \textbf{Activation Value Ranking (AVR):} AVR counts the cumulative activation value of each neuron using a batch of data, and ranks the neurons by the cumulative activation values.

\item \textbf{Activation Frequency Ranking (AFR):} AFR counts the times that neurons are activated (activation value $\alpha > 0$) using a batch of data as input, and ranks the neurons according to the number of times.

\item \textbf{Weight Value Ranking (WVR):} In each layer, WVR
counts the cumulative weight values of each neuron connection, and then sorts the cumulative weight values in descending order as the neuron importance ranking.

\item \textbf{Grad-CAM Ranking (GCR):} We design the Grad-CAM ranking (GCR), using the Grad-CAM technique to rank neurons in the convolutional layers. Grad-CAM does not rank the neurons in linear layers. Therefore we rank them in a random manner.

\item \textbf{Layer-wise relevance propagation Ranking (LRPR):} LRPR distributes the output correlation backward through the pre-trained model, and determines the contribution ranking of neurons to classification.

\end{itemize}

\begin{table*}[]
\centering
\caption{Evaluation results of EdgePro on different ranking strategies, i.e., RNR, AVR, AFR, WVR, GCR, LRPR. The values without and with brackets in the table represent $acc_{nl}$ and $acc_{nu}$.}
\label{tab:different ranking}
\resizebox{0.8\textwidth}{!}{%
\begin{tabular}{c|c|cccccc}
\hline
\multirow{2}{*}{\textbf{Dataset}}       & \multirow{2}{*}{\textbf{Model}}        & \multicolumn{6}{c}{\textbf{Method}}                                                                                                                                                                    \\ \cline{3-8}
                               &                               & \textbf{RNR}
                               & \textbf{AVR}
                               & \textbf{AFR}
                               & \textbf{WVR}
                               & \textbf{GCR}
                               & \textbf{LRPR}                          \\ \hline  \hline
\multirow{2}{*}{MNIST}         & \multirow{2}{*}{LeNet-5}      & 99.29\%                       & 99.60\%                       & 98.77\%                       & 99.64\%                       & 98.91\%                       & 99.11\%                       \\
                               &                               & \multicolumn{1}{l}{(10.01\%)} & \multicolumn{1}{l}{(10.70\%)} & \multicolumn{1}{l}{(9.80\%)}  & \multicolumn{1}{l}{(10.51\%)} & \multicolumn{1}{l}{(11.30\%)} & \multicolumn{1}{l}{(12.16\%)} \\ \hline
\multirow{2}{*}{CIFAR-10}      & \multirow{2}{*}{ResNet-18}    & 87.10\%                       & 87.01\%                       & 87.44\%                       & 86.97\%                       & 86.72\%                       & 87.02\%                       \\
                               &                               & \multicolumn{1}{l}{(11.71\%)} & \multicolumn{1}{l}{(8.20\%)}  & \multicolumn{1}{l}{(10.31\%)} & \multicolumn{1}{l}{(11.47\%)} & \multicolumn{1}{l}{(9.22\%)}  & \multicolumn{1}{l}{(9.27\%)}  \\ \hline
\multirow{2}{*}{CIFAR-100}     & \multirow{2}{*}{VGG-19}       & 73.52\%                       & 73.44\%                       & 72.81\%                       & 73.03\%                       & 72.49\%                       & 72.62\%                       \\
                               &                               & \multicolumn{1}{l}{(1.00\%)}  & \multicolumn{1}{l}{(0.69\%)}  & \multicolumn{1}{l}{(1.27\%)}  & \multicolumn{1}{l}{(0.93\%)}  & \multicolumn{1}{l}{(1.33\%)}  & \multicolumn{1}{l}{(1.03\%)}  \\ \hline
\multirow{2}{*}{Tiny-ImageNet} & \multirow{2}{*}{DenseNet-121} & 55.10\%                       & 55.12\%                       & 54.85\%                       & 55.23\%                       & 55.07\%                       & 54.59\%                       \\
                               &                               & \multicolumn{1}{l}{(0.35\%)}  & \multicolumn{1}{l}{(0.41\%)}  & \multicolumn{1}{l}{(0.40\%)}  & \multicolumn{1}{l}{(0.38\%)}  & \multicolumn{1}{l}{(0.50\%)}  & \multicolumn{1}{l}{(0.38\%)}  \\ \hline
\end{tabular}%
}
\end{table*}

\textbf{Results and Analysis.}
Table~\ref{tab:different ranking} shows the results of different ranking strategies.
Comparing the six ranking strategies, they all show good model protection with slight differences, for example, on the CIFAR-100 dataset, $acc_{nl}$ average reaches 72.99\% ($\pm$0.45\%) between different strategies.
All six different strategies have the effect of protecting models.
This means that the selection of authorization neurons does not depend on specific selection strategies, because the model can adapt to a small number of authorization neurons through training.
Therefore, it is appropriate to randomly select authorization neurons, which can fully ensure the unpredictability of EdgePro.

\subsection{Analysis of Parameter Sensitivity}
\label{sec 5.5}

\begin{figure}[ht]
\centering{
    \subfigure[MNIST]{
        \includegraphics[width=1.0\linewidth]{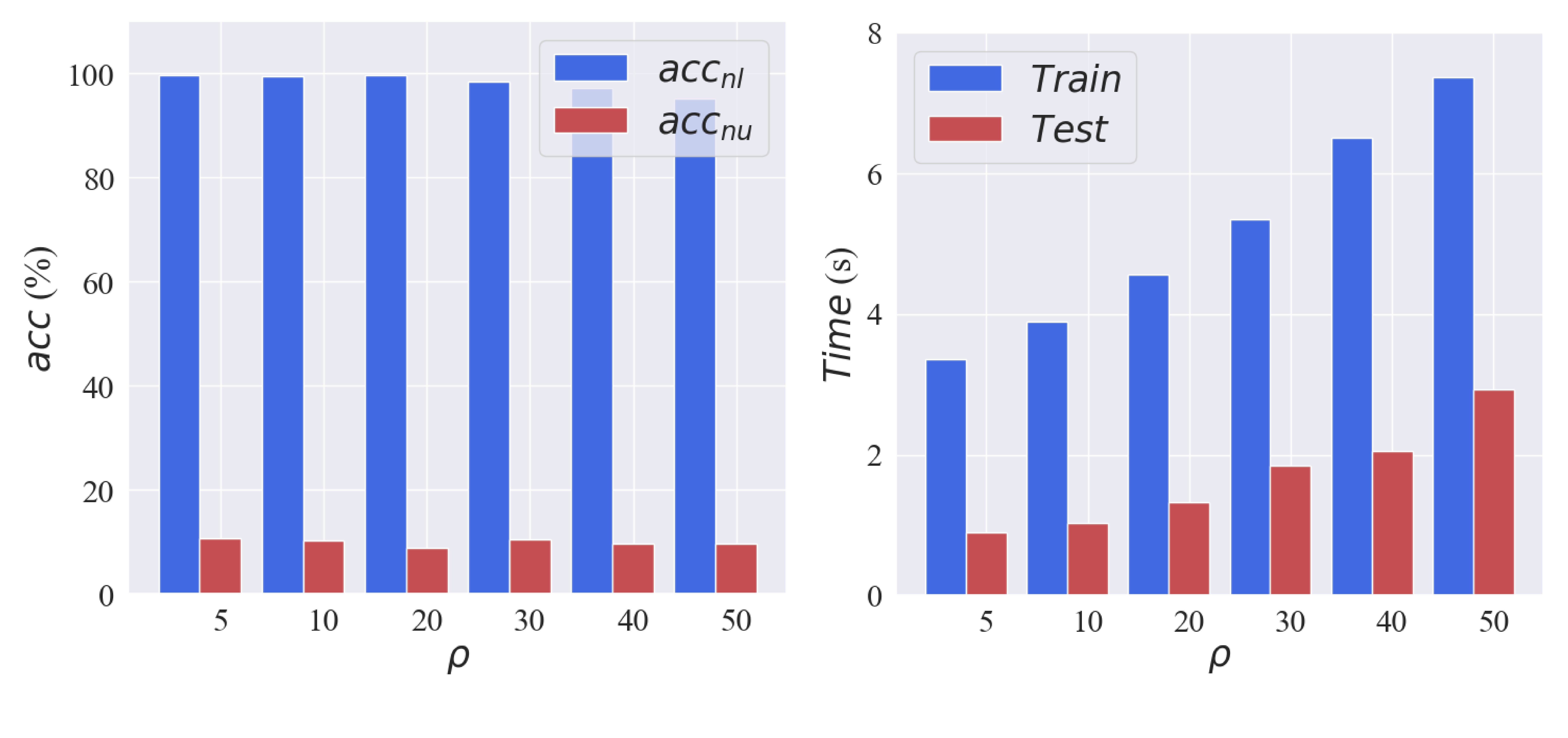} }
    \subfigure[CIFAR-10]{
        \includegraphics[width=1.0\linewidth]{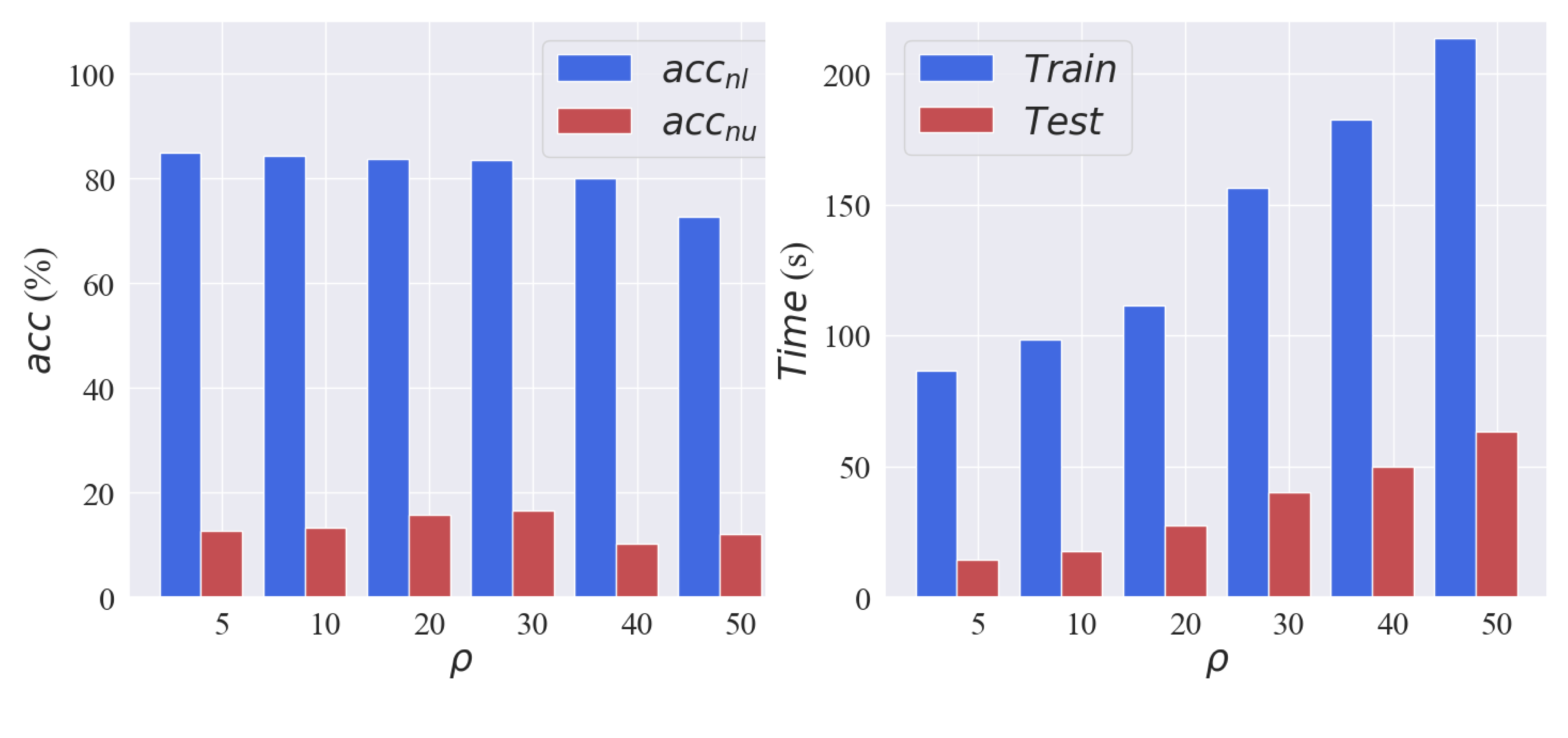} } }
    \\
\caption{The results of accuracy and time cost under different number of authorization neurons on MNIST and CIFAR-10.}
\label{The gamma of cifar}
\end{figure}

\subsubsection{The Effect of Authorization Neuron Number}
We provide an analysis of the effect about authorization neuron numbers by comparing model accuracy (both $acc_{nl}$ and $acc_{nu}$) and training time cost.

\textbf{Implementation Details.}
(1) We conduct experiments on the LeNet-5 model using 50,000 training data and 10,000 test data on the MNIST dataset and the ResNet-18 model using 50,000 training data and 10,000 test data on the CIFAR-10 dataset.
(2) We divide 6 different values: $\rho$=$\left \{5,10,20,30,40,50 \right \}$.
We record the $acc_{nl}$ and $acc_{nu}$ of EdgePro after running 20 epochs and compute the average training time in 1 epoch.
Finally, we record the testing time.

\textbf{Results and Analysis.}
Fig.~\ref{The gamma of cifar} shows the results.
As the authorization neuron ratio $\rho$ increases, the training cost of EdgePro also increases.
For instance, in Fig.~\ref{The gamma of cifar}(b) compared with $\rho=10$, when $\rho=50$, not only does $acc_{nl}$ drop by 13.8\%, but the training time per epoch is also increased by 3.6 times.
And in simple models, such as LeNet-5, the impact of $\rho$ on the accuracy is not obvious.
In fact, $\rho=5$ is also sufficient for the VGG-19 model with 50,782 neurons.
Therefore, it does not require many authorization neurons to achieve the protection purpose.
However, in consideration of the robustness against the reverse engineering attack, the neurons should not be too few, e.g., 1 or 2 neurons are shown to be vulnerable in Section~\ref{sec:robustness}.
We recommend 10\% of the neurons for models in comprehensive consideration of effectiveness, efficiency and robustness.

\begin{figure}[ht]
\centering{
    \subfigure[MNIST]{
        \includegraphics[width=0.7\linewidth]{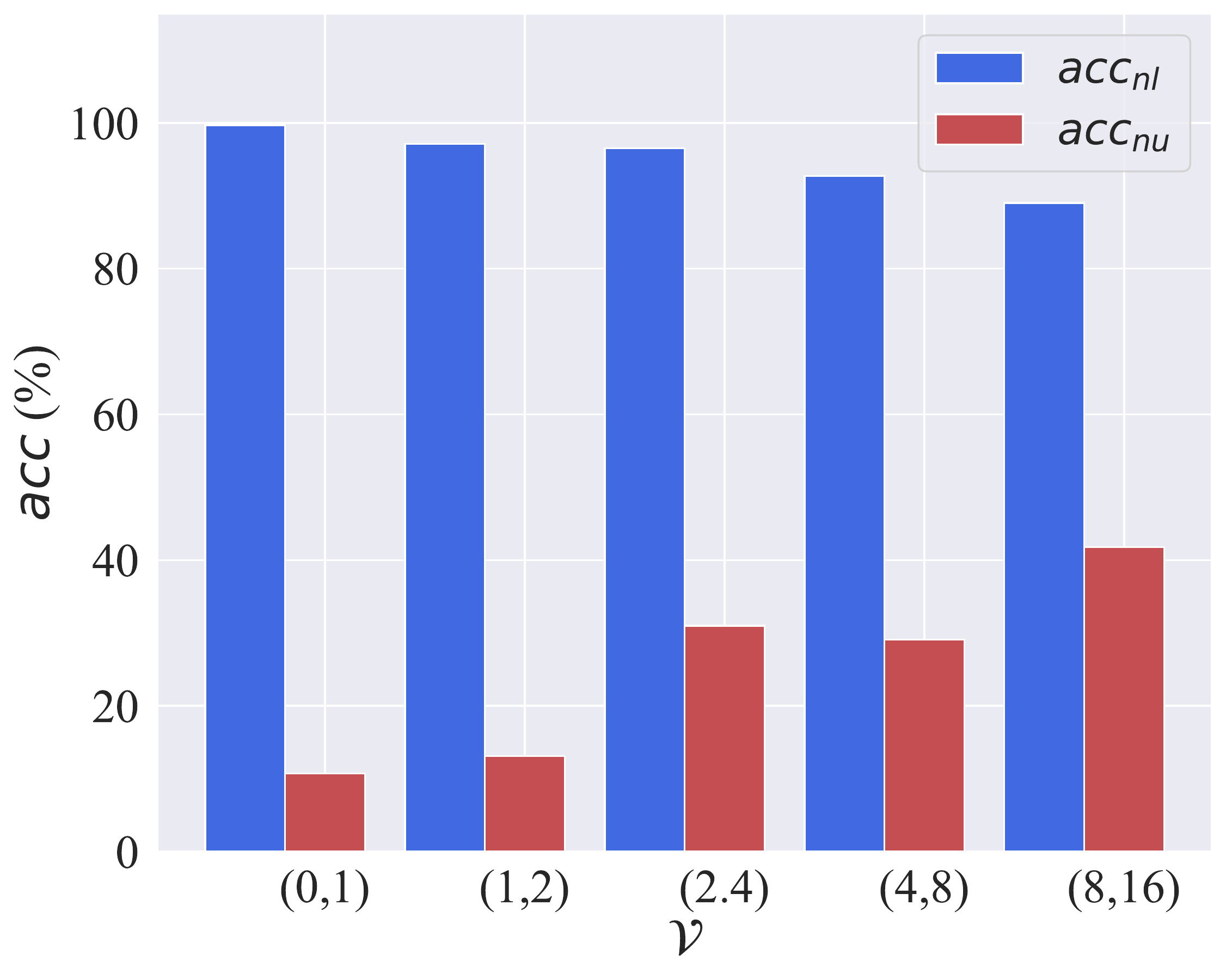} }
    \subfigure[CIFAR-10]{
        \includegraphics[width=0.7\linewidth]{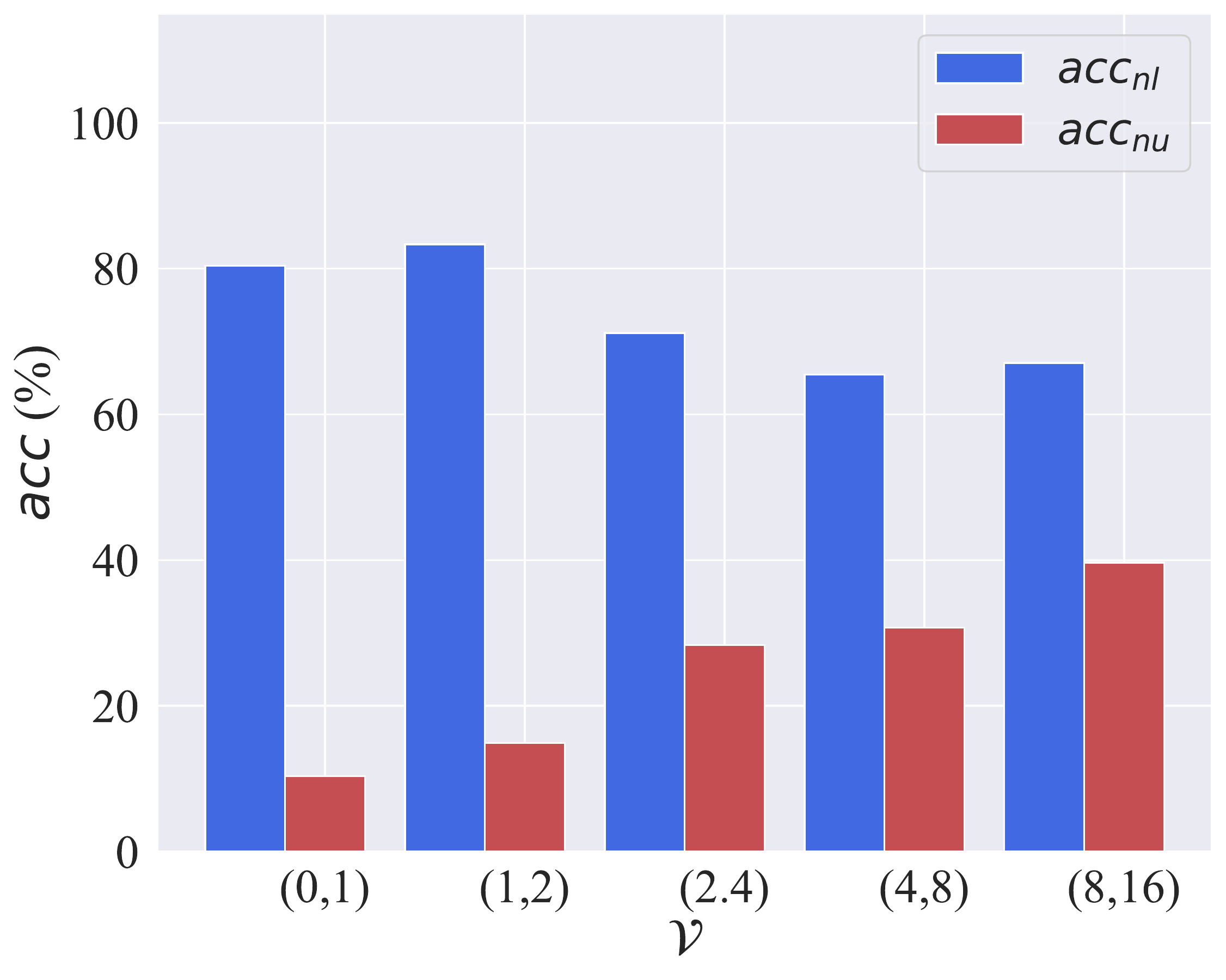} } }
    \\
\caption{The results of accuracy under different locking value size on LeNet-5 of MNIST and ResNet-18 of CIFAR-10.}
\label{The lambda of cifar}
\end{figure}

\subsubsection{The Effect of Locking Value}
We provide an analysis of the effect of locking value range $\mathcal{V}$ and perform the experiments on the LeNet-5 of MNIST and the ResNet-18 of CIFAR-10.

\textbf{Implementation Details.}
(1) We divide five locking value ranges: $\mathcal{V}$=$\left \{(0,1),(1,2),(2,4),(4,8),(8,16) \right \}$.
In each experiment, lock values will be randomly selected from the range.
(2) We record the $acc_{nl}$ and $acc_{nu}$ of EdgePro after running 20 epochs.

\textbf{Results and Analysis.}
Fig.~\ref{The lambda of cifar} shows the $acc_{nl}$ and $acc_{nu}$ of EdgePro trained model under different locking value ranges $\mathcal{V}$.
As $\mathcal{V}$ increases, the $acc_{nl}$ values demonstrate a decreasing trend.
When $\mathcal{V}=(8,16)$, $acc_{nl}$ is reduced by 16.3\% compared with $\mathcal{V}=(0,1)$.
Since EdgePro randomly selects authorization neurons, it may select some neurons that are irrelevant to the classification task. When the lock values of these irrelevant neurons are too large, the classification accuracy of the model will be affected, resulting in the reduction of $acc_{nl}$.
Meanwhile, too large locking values will also lead to excessive learning of the characteristics of authorization neurons during model training, resulting in the easy discovery of authorization neurons.
We consider $\mathcal{V}$=(0, 1) or $\mathcal{V}$=(1, 2) to be appropriate, at this time authorization neurons not only has little impact on model performance, but also are concealment.

\begin{figure}[ht]
\centering{
    \subfigure[MNIST]{
        \includegraphics[width=0.7\linewidth]{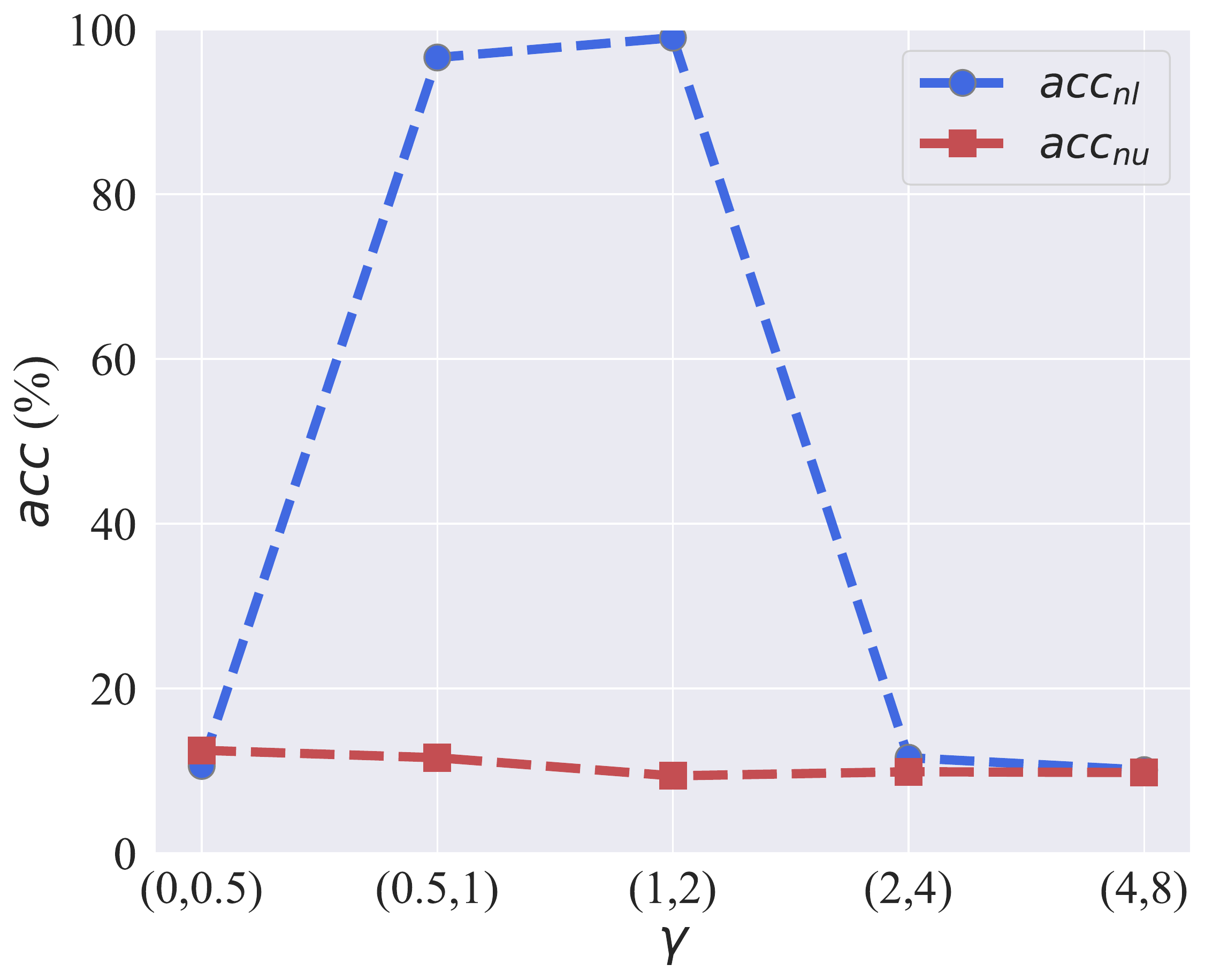} }
    \subfigure[CIFAR-10]{
        \includegraphics[width=0.7\linewidth]{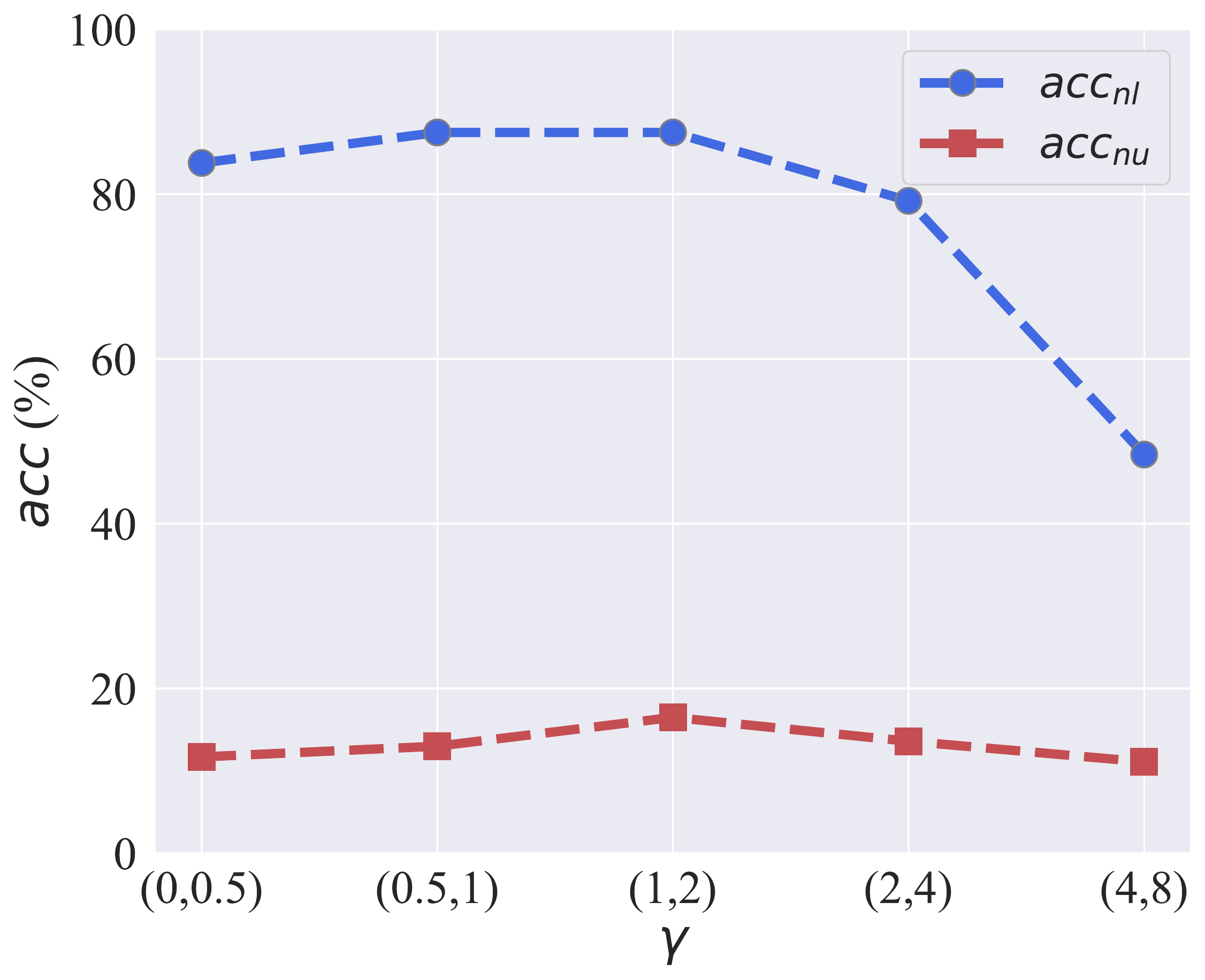} } }
    \\
\caption{The results of accuracy under different scale factors on LeNet-5 of MNIST and ResNet-18 of CIFAR-10.}
\label{The scale factor}
\end{figure}

\subsubsection{The Effect of Scale Factor}
We also provide an analysis of the effect of scale factors $\gamma$  by comparing the $acc_{nl}$ and $acc_{nu}$.

\textbf{Implementation Details.}
(1) We choose  MNIST with LeNet-5 and CIFAR-10 with ResNet-18 for experiments.
In view of the complex structure of ResNet, we multiply the output of each block by the scale factors instead of the output of each layer.

(2) We divide five scale factor ranges of different sizes: $\gamma \in$$\left \{(0,0.5),(0.5,1),(1,2),(2,4),(4,8) \right \}$.
As with the procedure for determining the locking value, we will also randomly select the scale factors from a range before starting the lock training.

\textbf{Results and Analysis.}
Fig.~\ref{The scale factor} shows the $acc_{nl}$ and $acc_{nu}$ of EdgePro.
It can be observed that inappropriate scale factors will affect the performance of the model.
For instance, on the MNIST dataset, both $\gamma \in$(0, 0.5) and $\gamma \in$(2, 4) will make the model unusable.
They scale the activation of neurons to an abnormal degree, resulting in the model can not being trained.
Additionally, compared with LeNet-5 scaling all layer outputs, only scaling block outputs in ResNet-18 will increase the tolerance of the model for scale factors.
This reflects that EdgePro does not need to add scale factors to all layers of the model.
It is sufficient to select some layers to add scale factors in a complex model.
Adding scale factors in too many layers will affect the performance of the model.
Therefore, for complex models, e.g., ResNet and DenseNet based models, EdgePro can add scale factors after each block instead of each layer.

\section{Future Work}
In this section, we will discuss the future work of EdgePro.
Theoretically, as long as there are neurons in the model, EdgePro ensures the security of the model through authorization at the neuron level.
In the experiments, we have explored and proved the effectiveness of EdgePro in image classification and node classification tasks.
In the future,  it would be interesting to extend EdgePro to more tasks, e.g., semantic segmentation, object detection/tracking, and text classification tasks.
In addition, generative adversarial networks also is one of our objectives.

\section{Conclusion}
We introduce a new protection method to the models on the edge devices, which is named EdgePro. Different from the existing methods, we embed the specific markers into part of the model neurons for the purpose of being light-weight. Only when the specific authorization neurons are locked to the locking values, the EdgePro trained model can work correctly. EdgePro replaces the encryption and storage of the entire model with the information
of authorization neurons. Our experiments show that the EdgePro is effective, light-weight, and robust against adaptive attacks including fine-tuning, reverse engineering and model pruning.

\bibliographystyle{IEEEtran}
\bibliography{EdgePro}

%

\begin{IEEEbiography}[{\includegraphics[width=0.8in,height=1.24in,clip,keepaspectratio]{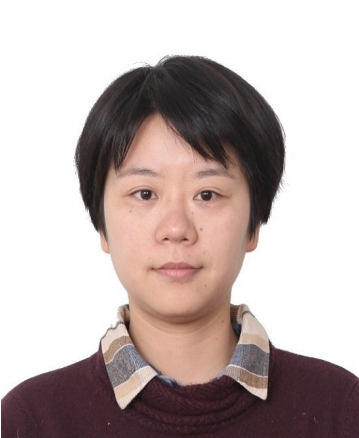}}]{Jinyin Chen} received BS and PhD degrees from Zhejiang University of Technology, Hangzhou, China, in 2004 and 2009, respectively. She studied evolutionary computing in Ashikaga Institute of Technology, Japan in 2005 and 2006. She is currently a Professor with the Zhejiang University of Technology, Hangzhou, China. Her research interests include artificial intelligence security, graph data mining and evolutionary computing.
\end{IEEEbiography}

\begin{IEEEbiography}[{\includegraphics[width=0.8in,height=1.24in,clip,keepaspectratio]{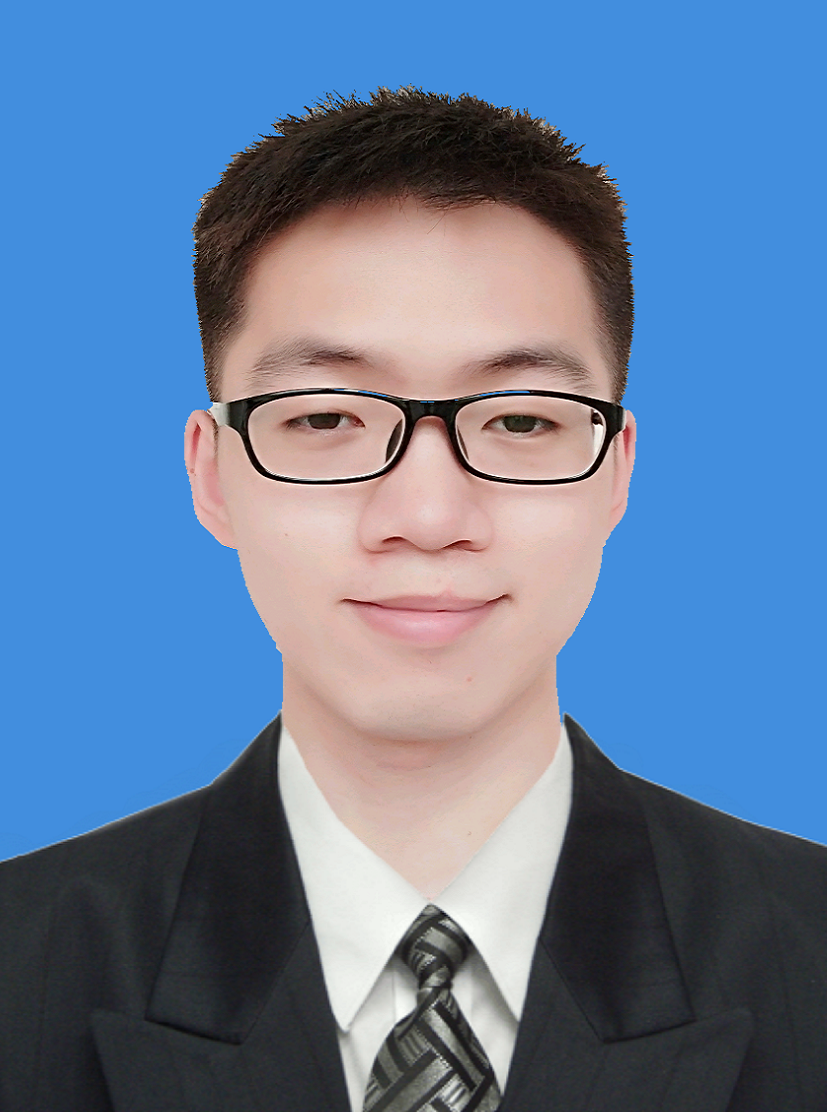}}]{Haibin Zheng} is a postdoc in the College of Computer Science and Technology at Zhejiang University of Technology.
He received B.S. and Ph.D. degrees both in Zhejiang University of Technology in 2017 and 2020, respectively.
His current research interests include Data-driven Security and Fairness, AI Security.
\end{IEEEbiography}

\begin{IEEEbiography}[{\includegraphics[width=0.8in,height=1.24in,clip,keepaspectratio]{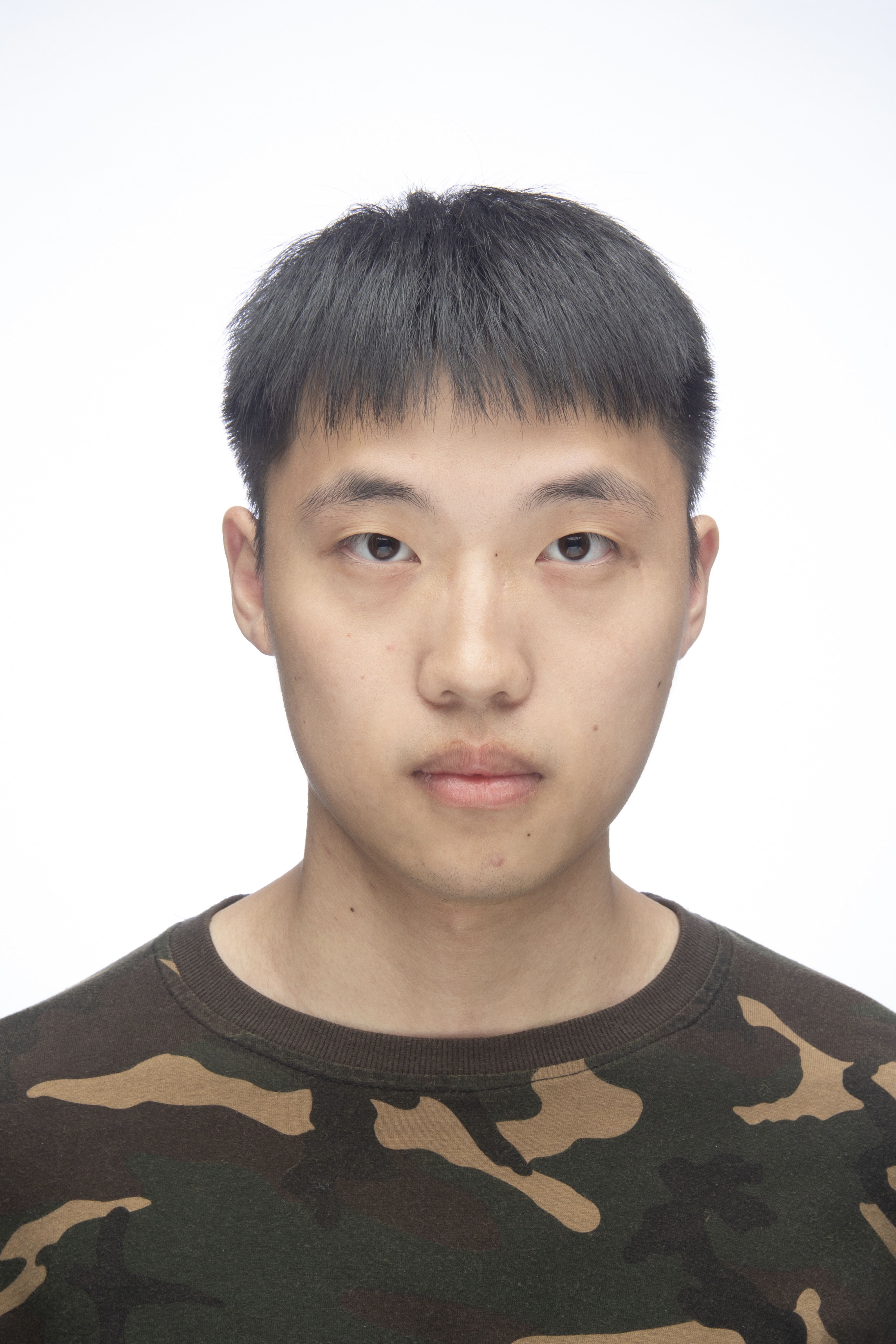}}]{Tao Liu} is currently pursuing the masters degree with the college of Information engineering, Zhejiang University of Technology. His research interests include federated learning and its applications, and artificial intelligence.
\end{IEEEbiography}

\begin{IEEEbiography}[{\includegraphics[width=0.8in,height=1.24in,clip,keepaspectratio]{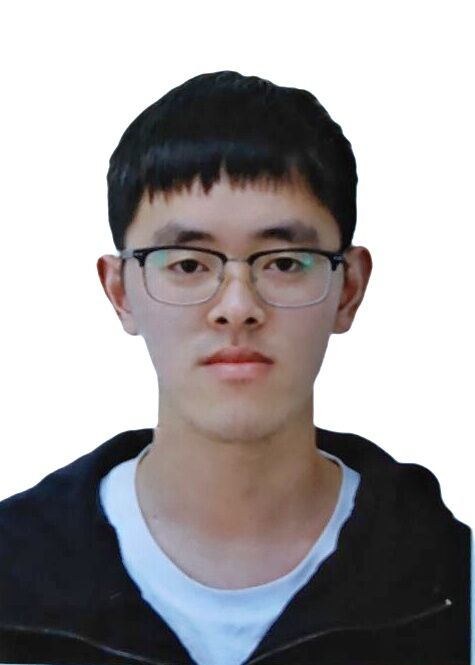}}]{Rongchang Li} is currently pursuing the masters degree with the college of Information engineering, Zhejiang University of Technology. His research interests include graph data mining and federated learning, and artificial intelligence.
\end{IEEEbiography}

\begin{IEEEbiography}[{\includegraphics[width=0.8in,height=1.24in,clip,keepaspectratio]{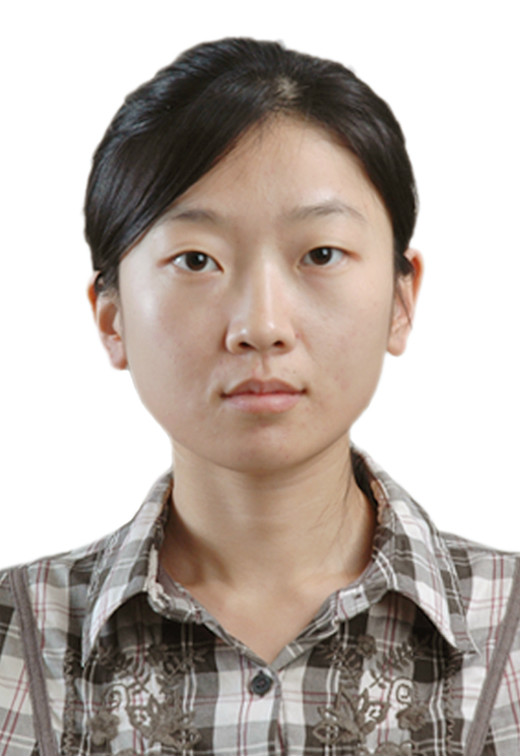}}]{Yao Cheng} is currently a senior researcher at Huawei International in Singapore. She received her Ph.D. degree in Computer Science and Technology from University of Chinese Academy of Sciences. Her research interests include security and privacy in deep learning systems, blockchain technology applications, Android framework vulnerability analysis, mobile application security analysis, and mobile malware detection.
\end{IEEEbiography}

\begin{IEEEbiography}[{\includegraphics[width=1.0in,height=1.55in,clip,keepaspectratio]{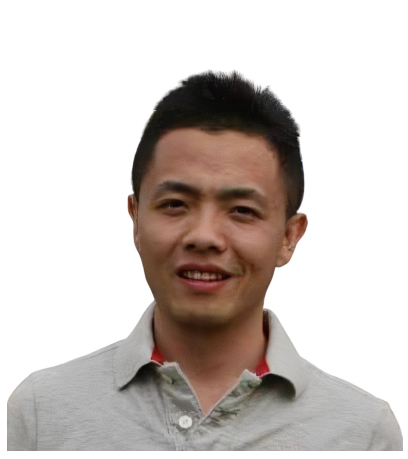}}]{Xuhong Zhang} received his Ph.D. in Computer Engineering from University of Central Florida in 2017. He received his B.S. Degree in Software Engineering from Harbin Institute of Technology in 2011 and received his M.S. degree in Computer Science from Georgia State University in 2013. He is an assistant professor of College of Control Science and Engineering at Zhejiang University. His research interests include distributed big data and AI systems, big data mining and analysis, data-driven security, AI and Security.
\end{IEEEbiography}

\begin{IEEEbiography}[{\includegraphics[width=0.9in,height=1.55in,clip,keepaspectratio]{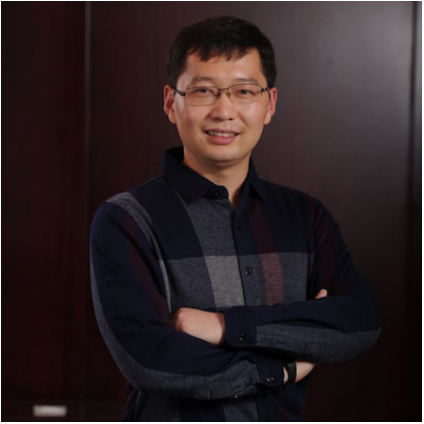}}]{Shouling Ji} is a ZJU 100-Young Professor in the College of Computer Science and Technology at Zhejiang University and a Research Faculty in the School of Electrical and Computer Engineering at Georgia Institute of Technology (Georgia Tech). He received a Ph.D. degree in Electrical and Computer Engineering from Georgia Institute of Technology, a Ph.D. degree in Computer Science from Georgia State University, and B.S. (with Honors) and M.S. degrees both in Computer Science from Heilongjiang University. His current research interests include Data-driven Security and Privacy, AI Security and Big Data Analytics.
\end{IEEEbiography}




\end{document}